\documentclass[reprint,aps,prd,twocolumn,showpacs,superscriptaddress,groupedaddress,amsmath,amssymb,]{revtex4}

\usepackage{graphicx}
\usepackage{dcolumn}
\usepackage{bm}
\usepackage{hyperref}
\usepackage{verbatim}

\def\be{\begin{equation}}
\def\bea{\begin{eqnarray}}
\def\eea{\end{eqnarray}}
\def\ee{\end{equation}}
\def\bi{\begin{itemize}}
\def\ei{\end{itemize}}
\def\cross{\times}

\def\bn{\begin{enumerate}}
\def\en{\end{enumerate}}

\def\be{\begin{equation}}
\def\ee{\end{equation}}
\def\bea{\begin{eqnarray}}
\def\eea{\end{eqnarray}}

\def\beq{\begin{equation}}
\def\eeq{\end{equation}}

\def\erfc{\mathrm{erfc}}

\begin{document}

\title{Method for all-sky searches of continuous gravitational wave signals using the frequency-Hough transform}
\author{Pia Astone$^a$, Alberto Colla$^{b,a}$, Sabrina D'Antonio$^c$, Sergio Frasca$^{b,a}$ and Cristiano Palomba$^a$ }
\affiliation{$^a$INFN, Sezione di Roma, P.le A. Moro, 2, I-00185 Roma, Italy\\$^b$Dip. di Fisica, Universita' di Roma ``Sapienza'', P.le A. Moro, 2, I-00185 Roma, Italy\\$^c$INFN, Sezione di Roma 2, Via della Ricerca Scientifica, 1, I-00133 Roma, Italy}

\begin{abstract} 
In this paper we present a hierarchical data analysis pipeline for all-sky searches of continuous gravitational wave signals, like those emitted by spinning neutron stars asymmetric with respect to the rotation axis, with unknown position, rotational frequency and spin-down. The core of the pipeline is an incoherent step based on a particularly efficient implementation of the Hough transform, that we call frequency-Hough, that maps the data time-frequency plane to the source frequency/spin-down plane for each fixed direction in the sky. Theoretical ROCs and sensitivity curves are computed and the dependency on various thresholds is discussed. A comparison of the sensitivity loss with respect to an ``optimal'' method is also presented. Several other novelties, with respect to other wide-parameter analysis pipelines, are also outlined. They concern, in particular, the construction of the grid in the parameter space, with over-resolution in frequency and parameter refinement, candidate selection and various data cleaning steps which are introduced to improve search sensitivity and rejection of false candidates.       

\end{abstract}


\maketitle

\section{\label{sec:intro}Introduction}

Continuous gravitational wave signals (CW) emitted by asymmetric rotating neutron stars are among the sources currently searched in the data of interferometric gravitational wave 
detectors. About $10^9$ neutron stars are expected to exist in the Galaxy. Of these, only about 2,400 have been detected through their electromagnetic emission, like pulsars. A fraction of the unseen population of neutron stars could in principle emit gravitational waves in the sensitivity band of detectors and it is therefore very important to develop efficient data analysis strategies to search the signals they emit.
Various mechanisms have been proposed that could allow for a time varying mass quadrupole in these stars, thus producing CW, like a residual crustal deformation or distortion induced by the inner magnetic field, see e.g. \cite{ref:owen2006} for a review.

Roughly speaking, CW searches are divided in {\it targeted}, when the source position and phase parameters are known with high accuracy, like in the case of known pulsars, 
and {\it all-sky} (also called {\it blind} or {\it wide-band}) in which those parameters are unknown and a wide portion of the parameter space is explored. In fact, also ``intermediate'' cases have been considered, like {\it narrow-band} searches \cite{ref:crab_nb}, \cite{ref:rome_nb} and {\it directed} searches \cite{ref:casa}. 

While {\it targeted} searches can be performed using optimal methods, based on matched filtering \cite{ref:knownpul2010}, \cite{ref:vela2011}, \cite{ref:knownpul2013}, this 
is practically impossible for {\it blind} searches, due to the huge number of points in the parameter space that must be typically explored. For this reason  
{\it hierarchical} procedures have been developed \cite{ref:power2}, \cite{ref:eh2013}, \cite{ref:houghsky}, \cite{c6c7} that allow a large reduction in the computational cost of the analysis at the price of a relatively small loss in sensitivity. Such
methods typically consists in dividing the whole data set in short pieces, each analyzed coherently, which are then combined incoherently, that is loosing the phase information.
Basically, three different kinds of incoherent steps have been proposed: the "stack-slide", the "PowerFlux", and the "Hough transform". The stack-slide procedure \cite{ref:stack1}, \cite{stack2} averages the normalized
power from the Fourier transform of 30-minute segments of the calibrated detector strain data. The PowerFlux schema \cite{ref:power1}, \cite{ref:power2} can be seen as a variation of the stack-slide, in which the power is
weighted before summing. The weights are chosen depending on the detector noise level and antenna pattern in such a way to maximize the signal-to-noise ratio (SNR). The Hough transform
method \cite{hough1},\cite{AdaHou},\cite{hough2} sums weighted counts, depending upon whether the equalized power in a Fourier transform bin exceeds a certain threshold and, depending on the specific algorithm implementation, other conditions are met. It is used for both ``short'' (of the order of the hour) and ``long'' (of the order of the day) time baseline searches. An ``optimal'', at least theoretically, incoherent method has also been studied \cite{ref:holg1}, \cite{ref:holg2} in the context of long time baseline searches. In fact, in some
cases different implementation of the same schema have been proposed. For instance, at least two "flavors" of the Hough transform method exist. The standard one \cite{hough1},\cite{AdaHou}, also used in the popular Einstein@Home hierarchical pipeline \cite{ref:eh2013}, in which for each fixed
value of the frequency and frequency derivative(s) a mapping between the time/frequency plane and the source position is done, and a newer one, called frequency-Hough (FH) \cite{HoughFFdot}, which is
based, for each fixed sky location, in a mapping between the time-frequency plane and the source frequency and spin-down plane. The FH transform has some important advantages with respect to the standard implementations both because a smaller sensitivity loss due to the digitizations involved in the procedure can be achieved without increasing the computational load and in terms of robustness with respect to disturbances.

In this paper we 
discuss a hierarchical procedure designed to effectively cope with the unavoidable problems raising when real data are used and putting attention to the practical implementative aspects of the analysis algorithms.
The core of the pipeline is the FH, which we fully characterize from a statistical point of view.
Moreover, we describe many other novel features with respect to other proposed hierarchical schemes, see e.g. \cite{ref:power2}, \cite{ref:eh2013}, regarding in particular the construction of the grid in the parameter space, the criteria for selecting candidates and the various cleaning steps applied to improve sensitivity and the capability of disregarding false candidates. 

The plan of the paper is the following. In Sec.\ref{signal} we describe the kind of gravitational wave signals we are looking for. In Sec.\ref{hier_scheme} we schematically present the whole scheme of the hierarchical procedure of which the FH constitutes the core. Details are given in the next sections. In Sec.\ref{sec:sfdb} the short FFT database is described. In Sec.\ref{sec:peakmap} the collection of time/frequency peaks, called {\it peakmap}, which is the input to the FH transform, is discussed. Sec.\ref{freqhough} is dedicated to the FH transform. In Sec.\ref{grid} we outline the construction of the {\it coarse} grid in the parameter space. In Sec.\ref{cand} we describe the criteria for
selecting candidates at the output of the FH. In Sec. \ref{refinedgrid} a {\it refined} analysis step around coarse candidates is presented. Candidate clustering and coincidences are discussed in Sec.\ref{clustandcoin}. Sec.\ref{verifollow} is about the final verification and follow-up step of the analysis procedure. Sec.\ref{sens} is devoted to the theoretical computation of ROC curves and search sensitivity. Next section, Sec.\ref{pmclean}, is about the various cleaning steps that are applied in order to discard disturbances. Finally, in Sec.\ref{conc} conclusions and future prospects are discussed. Some mathematical and implementative details are given in the Appendix.

\section{\label{signal}Continuous gravitational wave signals from spinning neutron stars}

The expected quadrupolar gravitational-wave signal at the detector from a non-axisymmetric neutron star steadily spinning about one of its principal axis is 
at twice the rotation frequency $f_{rot}$, with a strain of \cite{ref:fivevect}
\be
h(t)=H_0(H_+A^++H_{\times}A^{\times})e^{\jmath \left(\omega(t)t+\Phi_0\right)}
\label{eq:hoft}
\ee
where taking the real part is understood. The signal frequency and phase at time $t_0$ are, respectively, $f_0=\frac{\omega(t_0)}{2\pi}=2f_{rot}(t_0)$ and $\Phi_0$. The two complex amplitudes $H_+$ and $H_{\times}$ are given respectively by
\begin{equation}
H_+=\frac{\cos{2\psi}-\jmath \eta \sin{2\psi}}{\sqrt{1+\eta^2}}
\label{eq:Hp}
\end{equation}
\begin{equation}
H_{\times}=\frac{\sin{2\psi}+\jmath \eta \cos{2\psi}}{\sqrt{1+\eta^2}}
\label{eq:Hc}
\end{equation}
in which $\eta$ is the ratio of the polarization ellipse semi-minor to semi-major axis and the polarization angle $\psi$ defines the direction of the major axis with respect to the celestial parallel of the source (counterclockwise). The parameter $\eta$ varies in the range $[-1,1]$, where $\eta=0$ for a linearly polarized wave and $\eta=\pm 1$ for a circularly polarized wave ($\eta=1$ if the circular rotation is counterclockwise). 
The functions $A^+$ and $A^{\times}$ describe  the detector response as a function of time and are given by
\begin{align}
A^+ =  & a_0+a_{1c} \cos{\Omega_{\oplus} t}+a_{1s} \sin{\Omega_{\oplus} t} +a_{2c} \cos{2\Omega_{\oplus} t}+\\ \nonumber
& a_{2s} \sin{2\Omega_{\oplus} t}  
\label{eq:Ap}
\end{align}
\begin{align}
A^{\times} =  &b_{1c} \cos{\Omega_{\oplus} t}+b_{1s} \sin{\Omega_{\oplus} t} +b_{2c} \cos{2\Omega_{\oplus} t}+\\ \nonumber 
& b_{2s} \sin{2\Omega_{\oplus} t}
\label{eq:Am}
\end{align}
where $\Omega_{\oplus}$ is the Earth sidereal angular frequency and with the coefficients depending on the source position and detector position and orientation on the Earth \cite{ref:fivevect}. 

As discussed in \cite{ref:vela_vsr2} the strain described by Eq.(\ref{eq:hoft}) is equivalent to the standard expression, see e.g. \cite{ref:jks}
\begin{align}
h(t) =  &\frac{1}{2}F_+(t, \psi)h_0(1+\cos{}^2\iota)\cos{\Phi(t)} \\ \nonumber 
& + F_{\times}(t, \psi)h_0\cos{\iota}\sin{\Phi(t)}
\label{eq:hclass}
\end{align}
Here $F_+,~F_{\times}$ are the ``classical'' beam-pattern functions, $\iota$ is the angle between the star rotation axis and the line of sight; the amplitude 
\begin{equation}\label{eq:h0}
h_0=\frac{4\pi^2G}{c^4}\frac{I_{zz}\varepsilon f^2_0}{d}
\end{equation}
depends on $I_{zz}$, which is the star moment of inertia with respect to the 
principal axis aligned with the rotation axis, on $\varepsilon=\frac{I_{xx}-I_{yy}}{I_{zz}}$ which is the equatorial ellipticity expressed in terms of principal moments of inertia and on $d$, which is the source distance. 
While estimations of the maximum braking strain that a neutron star crust can sustain have been
done and strongly depend on its structure and equation of state (see, e.g., \cite{ref:horo}, \cite{ref:john_owen}, \cite{ref:ciolfi}), the actual ellipticity is largely unknown.  
The relation between $H_0$ and $h_0$ is given by
\begin{equation}
H_0=h_0 \sqrt{\frac{1+6 \cos^2 \iota+ \cos^4 \iota}{4}}
\end{equation} 
while
\begin{equation}
\eta=-\frac{2 \cos \iota}{1+\cos^2 \iota} 
\end{equation}

In  Eq.(\ref{eq:hoft}) the signal angular frequency $\omega(t)$ is a function of time, and then the signal phase 
\begin{equation}
\Phi(t)=\int_{t_0}^t \omega(t')dt'
\label{eq:phit}
\end{equation} 
is not that of a simple monochromatic signal and depends on both the intrinsic rotational frequency and frequency derivatives of 
the neutron star and on Doppler and propagation effects. These effects include relativistic modulations 
caused by the Earth's orbital and rotational motion \footnote{For a source in a binary system also the binary orbital motion must be taken into account.} and the presence of massive bodies in the solar system close 
to the line-of-sight to the pulsar.
The received Doppler-shifted frequency $f(t)$ is related to the emitted frequency $f_0(t)$ by the well-known relation (valid in the non-relativistic approximation)
\begin{equation}
f(t)=\frac{1}{2\pi}\frac{d\Phi(t)}{dt}= f_0(t) \left(1+\frac{\vec{v}
\cdot \hat{n}}{c}\right),
\label{eq:fdopp}
\end{equation}
where $\vec{v}=\vec{v}_{orb}+\vec{v}_{rot}$ is the detector velocity with respect to the Solar system barycenter (SSB), sum of the Earth orbital velocity around the Sun, $\vec{v}_{orb}$, and of the Earth 
rotation velocity, $\vec{v}_{rot}$, while $\hat{n}$ is the versor identifying the source position and $c$ is the light velocity. In terms of equatorial
coordinates $(\alpha, \delta)$, the components of the versor $\hat{n}$ are (cos{$\alpha$}cos{$\delta$},~sin{$\alpha$}sin{$\delta$},~sin{$\delta$}).   

The intrinsic signal frequency $f_0(t)$ slowly decreases in time due to the source spin-down, associated to the rotational energy loss following emission of electromagnetic and/or gravitational radiation. The spin-down can be described through a series expansion
\begin{equation}
f_0(t)=f_0+\dot{f}_0(t-t_0)+\frac{\ddot{f}_0}{2}(t-t_0)^2+...
\label{eq:sd}
\end{equation}
In general a CW depends then on 3+$s$ parameters: position, frequency and $s$ spin-down parameters.

!\section{\label{hier_scheme}Scheme of the hierarchical procedure}
All-sky searches cannot be afforded with a completely coherent method, due to the huge dimension of the
parameter space which poses challenging computational problems \cite{ref:stack1}, \cite{sergioqui}. Moreover, a completely coherent 
search would not be robust towards unpredictable phase variations of the signal 
during the observation time. For these reasons hierarchical schemes have been developed.
The hierarchical scheme we present starts from the detector calibrated data. The first step consists in constructing a {\it short FFT database} (SFDB)
\cite{SFDB} where each FFT is built from a data chunk of duration, called {\it coherence time}, short enough such that if a signal is present its frequency, which is modified by the Doppler and
spin-down described in previous section, remains within a frequency bin. The FFT duration is then a function of the search frequency, with longer
FFTs allowed at lower frequencies. From the SFDB we create a time-frequency map, called {\it peakmap}  \cite{AdaHou} \cite{Cleaning}, obtained selecting the
most significant peaks on equalized periodograms. The peakmap is the input of the incoherent step, based on the FH
transform \cite{HoughFFdot}. In the FH transform we take into account also noise slow non-stationarity and the varying detector sensitivity caused by
the time-dependent radiation pattern. The most significant candidates are selected at this stage using a coarse grid in the parameter space and an effective way to avoid blinding by particularly disturbed frequency bands.
For each coarse candidate a refined search is run again on the 
neighborhood of the candidate parameters and the final {\it first level} refined candidates are selected. Candidates are then {\it clustered}, grouping toghether those occupying nearby points in the parameter space. In order to significantly reduce the false alarm
probability, {\it coincidences} are done among clusters of candidates obtained in the analysis of different data sets (of the same detector or of different detectors). Over
coincident candidates, after a {\it verification} step, a {\it follow-up}, with a longer coherence time, is applied. 

The choice of using ``short'' FFTs is similar to the one done, for instance, in the PowerFlux pipeline \cite{ref:power1} and the standard Hough transform search described in \cite{ref:houghsky}. Another popular hierarchical pipeline, Einstein@Home \cite{ref:eh2013}, uses the F-statistic \cite{ref:jks},\cite{ref:fstatnet} coherently computed on about 1 day data segments from multiple detectors, followed by the standard Hough transform as incoherent step. At least theoretically this clearly gives a gain in sensitivity. On the other hand, this choice is less robust against unforseen GW signal frequency modulations on time scales smaller than about 1 day while, on the contrary, splitting the data in a larger number of shorter segments is also more robust against disturbances in one segment. We expect that the use, in the pipeline we have developed, of the FH transform, of the refinement only around coarse candidates, and various aggressive cleaning steps allows to significantly improve the detection efficiency and to partially compensate the shorter FFT length against a lower computational load. By the way, each pipeline uses its optimization tricks and its cleaning procedures so that a comparison of the performance of our analysis pipeline with other methods would be really meaningful only if real data were used but is outside the scope of this paper.     

\begin{figure*} [hbtp]  
\includegraphics[width=10cm]{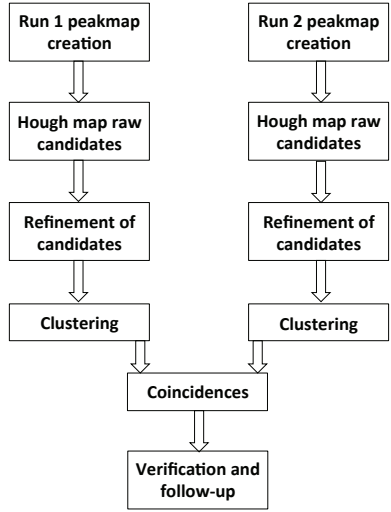}
\caption{Scheme of the hierarchical pipeline. See text for a description of the various blocks.}  
\label{fig:HierScheme}
\end{figure*}

In Fig.(\ref{fig:HierScheme}) the main steps of the hierarchical procedure are shown and will be discussed in following sections, briefly recalling those already
presented in previous papers, and focusing attention on new choices and improvements with the aim of presenting a coherent and unified view of the full analysis pipeline. 


\section{\label{sec:sfdb}Short FFT Data Base}
The SFDB construction and characteristics 
have been described in \cite{SFDB}. 
Shortly, it is  a collection of FFTs 
obtained from detector calibrated data 
divided into interlaced (by half) chunks of proper 
time duration, each windowed in order to reduce the dispersion 
of power due to their finite length.
The time duration $T_{FFT}$ of each FFT is chosen using the criterium that if a signal is present the frequency spread due to the 
Doppler effect is smaller than a frequency bin during the time $T_{FFT}$. It can be shown \cite{sergioqui} that the maximum FFT duration is given by $\sim \frac{1.1 \times 10^5}{\sqrt{f_{max}}}$ seconds where $f_{max}$ is the maximum frequency of the FFT in Hertz. 
A data cleaning procedure, described in \cite{Cleaning}, is applied in time-domain when constructing the data base, see Sec.\ref{pmclean}, in such a way to not throw away available data and at the same time to improve 
the sensitivity through the identification and removal of large 
and short time duration disturbances.

Given that the maximum FFT duration is a function of the frequency, the SFDB is divided into blocks covering different frequency ranges, with FFT length depending on the maximum block frequency.  
In table \ref{tab:parSFDB} we give a possible organization of the SFDB in terms of frequency bands, their corresponding sampling times and FFT time durations.
Table \ref{tab:parSFDB2} shows another possible choice, 
aimed to contain the computational cost, obviously with a consequent loss of sensitivity at high frequency. In the remaining of the paper most of the examples and plots involving the SFDB refer to this choice.

\begin{table}
\begin{center}
\begin{tabular}{|c|c|c|c|c|}
\hline
 $B $ [Hz] & $ T_{FFT}$ [s]  & $ \delta t$ [s]& $\delta f$[Hz] & $N_{f}$ \\
\hline
\hline
&&&& \\
$512-2048$ & 1024  & $2.44\cdot 10^{-4}$  & $9.77 \cdot 10^{-4}$  & $1.57 \cdot 10^6$ \\
&&(1/4096)&(1/1024)& \\
&&&& \\
$128- 512$ & 4096  & $ 9.77\cdot 10^{-4}$  & $2.44\cdot 10^{-4}$ & $1.57 \cdot 10^6$ \\
&&(1/1024)&(1/4096)& \\
&&&& \\
$32- 128$ & 8192  & $ 3.91\cdot 10^{-3} $  & $1.22 \cdot 10^{-4}$ & $ 7.86 \cdot 10^5$ \\
&&(1/256)&(1/8192)& \\
&&&& \\
$10- 32$ & 16384  & $ 1.56 \cdot 10^{-2} $ & $6.10 \cdot 10^{-5}$ & $ 3.60\cdot 10^5$ \\
&&(1/64)&(1/16384)& \\
&&&& \\
\hline
\end{tabular}
\caption{The table shows a possible organization of the short FFT data base, using four frequency bands. 
$T_{FFT}$ is the time duration of each FFT, $\delta t$ is the sampling time, $\delta f$ is the frequency resolution of the FFT and $N_f$ the number of frequency bins.}
\label{tab:parSFDB}
\end{center}
\end{table}

\begin{table}
\begin{center}
\begin{tabular}{|c|c|c|c|c|}
\hline
 $B $ [Hz] & $ T_{FFT}$ [s]  & $ \delta t$ [s]& $\delta f$[Hz] & $N_{f}$ \\
\hline
\hline
&&&& \\
$128-2048$ & 1024  & $2.44\cdot 10^{-4}$   & $9.77 \cdot 10^{-4}$  & $1.97 \cdot 10^6$ \\
&&(1/4096)&(1/1024)& \\
&&&& \\
$10-128$ & 8192  & $ 3.91 \cdot 10^{-3} $  & $1.22 \cdot 10^{-5}$ & $ 9.67\cdot 10^5$ \\
&&(1/256)&(1/8192)& \\
&&&& \\
\hline
\end{tabular}
\caption{Another possible organization of the SFDB, which reduces the computational load by penalizing a bit the sensitivity at high frequency.}
\label{tab:parSFDB2}
\end{center}
\end{table}
For each FFT also a lower resolution auto-regressive estimation of the average spectrum, called {\it very short FFT} is computed and stored in the database.

\section{\label{sec:peakmap}Peakmap}
For each of the N FFTs in the SFDB we compute the periodogram, $S_{p;i}(f)$, $i=1,...N$, i.e. the square modulus of the FFT, and then the ratio between the periodogram and the auto-regressive average spectrum estimation, $S_{AR;i}(f)$:
\be
R(i,j)=\frac{S_{P;i}(f)}{S_{AR;i}(f)};~~i=1,...N
\label{pm}
\ee
where the ratio is computed frequency bin by frequency bin and $j$ runs over the frequency bins of the $i$th FFT. The function $R(i,j)$ is compared to a threshold $\theta$ and the frequency bins which are above the threshold {\it and} are local maxima are selected. Each pair made of a selected frequency bin and of the initial time of the corresponding FFT is a peak. Note that, differently from what is done in the ``Stack-slide'' \cite{ref:stack1} and ``PowerFlux'' schema \cite{ref:power1}, the peak amplitude is not taken into account. The collection of all the peaks, considering all the FFTs of the SFDB forms the {\it peakmap}. Selecting peaks which are above the threshold and also local maxima has some important advantages with respect to the choice done e.g. in \cite{hough1}, \cite{ref:eh2013}, where only the first condition is considered: less sensitivity to spectral disturbances (i.e. better robustness) and a significantly lower computational cost of the analysis, because the number of peaks is smaller. On the other hand, as we will see, this choice implies also a very small theoretical sensitivity loss.

The Hough transform is computed starting from the peak-map. If a
peak is selected at the level of the peakmap it will contribute to the Hough number
count, even if it is due to noise. On the other hand, if a signal peak is missed at
the peakmap level it will not contribute to the Hough map. Let us indicate with
$p_0=P(\theta;0)$ the probability of selecting a noise peak above the threshold
$\theta$ in the peakmap and with $p_{\lambda}=P(\theta;\lambda)$ the probability
when a signal with spectral amplitude (in units of equalized noise) $\lambda$ is
present. This is defined as
\be
\lambda=\frac{4|\tilde{h}(f)|^2}{T_{FFT}S_n(f)}
\label{lambdadef}
\ee
where $\tilde{h}(f)=\int_{-\infty}^{+\infty} h(t)e^{-j2\pi f t}dt$ is the Fourier transform of the signal $h(t)$ and $S_n(f)$ is the detector uni-lateral noise spectral density. 
In practice, $p_0$ is the false alarm probability for noise peak selection, while
$1-p_{\lambda}$ is the false dismissal probability for signal peak selection. 
In case of gaussian noise the probability distribution of the power in each bin of a periodogram is exponential with mean value equal to the standard deviation. For the peakmap, given that dividing the periodogram by the auto-regressive average spectrum estimation we are in fact making an equalization, the probability distribution is still exponential with mean value and standard deviation equal to 1.  
Then, $p_0$ can be computed observing that the probability of having in 
the $j$th frequency bin of the $i$th FFT a value of the ratio $R(i,j)$ between $x$ and $x+dx$ is 
$e^{-x}dx$. The probability that that given value is also a local maxima is easily
obtained multiplying by the probability that the two neighboring bins have a smaller
value, that is $(1-e^{-x})^2$. Then the probability of having a local maxima above a
threshold $\theta$ is 
\be
p_0=\int_{\theta}^{+\infty} 
e^{-x}(1-e^{-x})^2dx=e^{-\theta}-e^{-2\theta}+\frac{1}{3}e^{-3\theta}
\label{eq:p0}
\ee
The probability of having $n$ peaks in a peakmap is a binomial with expectation value $mp_0$ and standard deviation $\sqrt{mp_0(1-p_0)}$, being $m=N\cdot N_f$ the total number of bins in the peakmap.
 
In presence of a signal with spectral amplitude $\lambda$ the probability density of the spectrum is a
normalized non-central $\chi ^2$ with 2 degrees of freedom and non-centrality parameter
$\lambda$:
\begin{equation}
p(x;\lambda)= e^{(-x-\frac{\lambda}{2})}I_0(\sqrt{2x \lambda}) 
\label{noncenchi2}
\end{equation}
where $I_0$ is the modified Bessel function of zeroth order, which has 
mean value $\bar{x} = 1+\frac{\lambda}{2}$ and variance
$\sigma^2_x = 1+\lambda $. 
For small signals, i.e. $\lambda \ll x$ we have that
\begin{align}
e^{(-x-\frac{\lambda}{2})}\approx e^{-x}(1-\frac{\lambda}{2}) \notag \\
I_0(\sqrt{2x \lambda})\approx 1+\frac{2x\lambda}{4}
\end{align}
then
\be
p(x;\lambda)\approx e^{-x}\left(1-\frac{\lambda}{2}+\frac{\lambda}{2}x\right)
\ee
The probability of selecting a local maxima above a threshold can be computed as
before:
\be
p_{\lambda}=\int_{\theta}^{+\infty}p(x;\lambda)\left(\int_0^x
p(x';\lambda)dx'\right)^2dx
\ee
The inner integral is equal to $1-e^{-x}-\frac{\lambda}{2}xe^{-x}$ and the final
result, by keeping terms only up to $o(\lambda)$ is 
\be
p_{\lambda}\approx p_0+\frac{\lambda}{2}\theta \left(e^{-\theta}-2
e^{-2\theta}+e^{-3\theta}\right)
\label{eq:plambda_approx}
\ee
The choice of the threshold $\theta$, which impact on the search sensitivity and computational weight of the analysis, will be discussed in Sec.\ref{sens}.

\section{\label{freqhough}Frequency-Hough transform}
The Hough transform is a processing techniques for robust pattern extraction mainly from digital images. In CW searches it is used to map points in the time/frequency plane which follow the pattern expected from a signal into the signal parameter space. As mentioned in Sec.\ref{sec:intro}, various implementations of the Hough transform exist.   
Here we summarize the basic concepts of the FH transform, first introduced in \cite{HoughFFdot}.
We assume the second order spin-down can be neglected. 
As will be shown in Sec.\ref{grid} this corresponds to a constrain on
the so-called minimum spin down age.
The FH consists in a linear mapping between the {\it detector time/source Doppler corrected frequency} 
plane into the {\it source intrinsic frequency/spin-down} plane.
If $f$ is the signal frequency at the detector
(Doppler corrected for a given sky direction), 
$f_0$ the source intrinsic frequency at time $t_0$, $\dot{f}_0$ the first 
spin-down parameter and $t$ the time at the detector,
we have that
\begin{equation}
f=f_0+ \dot{f}_0\,(t-t_0)
\end{equation}
Hence
\begin{equation}
\dot{f}_0= -\frac{f_0}{t-t_0} + \frac{f}{t-t_0}
\label{eq:uno}
\end{equation}
The input plane is obtained from the original peakmap
by correcting it for the Doppler shift due to the Earth motion
for each point in the sky grid we analyze.
As, by construction, each FFT in the SFDB is short enough that the signal power is confined within a single frequency bin, see Sec. \ref{sec:sfdb}, the removal of the Doppler effect from the original peakmap consists in a simple {\it shifting} of the 
peakmap bins. Each point in the input plane $(t-t_0,f)$ is transformed into  a straight line in the
$(f_0,\dot{f}_0)$  Hough plane, with slope $-1/(t-t_0)$. 
In fact, by taking into account the width $\delta f_H$ of the frequency bins in the 
input plane, see Eq.(\ref{eq:freqres}), each peak is transformed into a stripe delimited by 
two parallel straight lines and covering a range of spin-down values given by
\begin{equation}
 -\frac{f_0}{t-t_0} +\frac{f - \delta f_H/2}{t-t_0}
< \dot{f}_0 < -\frac{f_0}{t-t_0} +\frac{f+ \delta f_H/2}{t-t_0}\label{eq:due}
\end{equation}
In each bin of the frequency/spin-down plane touched by a stripe the number count is increased by one. For each fixed direction in the sky, the set of number counts in the frequency and spin-down bins constitutes an Hough map (or Hough histogram). The number count $n$ in a given bin can be seen as the sum of binary counts $n_i$, which takes value 0 or 1:
\begin{equation}
n=\sum_{i=1}^{N}n_i
\label{ncount}
\end{equation}
The probability distribution of the Hough map is then binomial, i.e. the probability of
having a number count $n$ in a given pixel of a map built starting from $N$ FFTs is the same for both the classical Hough and the FH and is given by \cite{hough1} 
\be
P_n(\theta;\lambda)={N\choose n} \eta^n\left(1-\eta\right)^{N-n}
\label{eq:phmsig}
\ee
where $\eta=p_0$ when no signal is present and $\eta=p_{\lambda}$ when a signal is present.
The mean and variance of the number count are respectively
\begin{align}
\mu=N\eta \notag \\
\sigma^2=N\eta(1-\eta)
\label{houghmean}
\end{align}

In presence of a signal strong enough the stripes corresponding to the various input peaks, when the correct source position is considered, intersect in the transformed plane identifying the intrinsic frequency and spin-down of the source.
In practice, we are interested in those bins of the Hough map where the number count is high with respect to the average value. The slope of these stripes depends on 
the choice of the reference time $t_0$. By putting the reference time in the
middle of the observation time it is possible to see that the signal affects the smallest possible number of pixels in the parameter space, thus reducing the contamination of nearby pixels. 

The FH transform, and the specific way in which it is implemented, presents some relevant differences with respect to typical implementations of the standard Hough transform \cite{hough1,AdaHou},  
where the transformation is between the time-frequency peakmap and the celestial sphere, which is not computationally light due to the non linearity of the mapping. This has some important consequences. First of all, 
to reduce the computational effort ``look-up tables'' 
are used in the standard Hough to speed-up the mapping between the input and transformed planes and this introduces further digitization errors in addition to those intrinsic to the Hough mapping and due to the finite resolution.
Again to reduce the computational load of the analysis, fast algorithms have been developed,
which require the use of a rectangular grid in the sky. Compared to the
grid actually used for the FH, see Sec. \ref{grid}, the rectangular one has
over-resolution in some regions of the sky, 
which increases the number of points in the parameter space. Moreover
the use of the sky as the space to spot candidates is very 
prone to artifacts: some regions are always ``privileged'', that is 
they have a higher number of candidates with respect to the expectation. 
On the other hand, in the FH enhancing the frequency resolution does not 
cost from a computational point of view but reduces the 
discretization loss due to the finite size of the bins. The effect of over-resolution in frequency has been studied with simulated signals in \cite{HoughFFdot} and the main result is that using an over-resolution factor in frequency of 10 also affects the efficiency loss associated with the sky grid and gives an overall efficiency loss of about 13$\%$ with respect to about 25$\%$ if the over-resolution is not used, with a ratio of the sensitivity losses of $\sim 0.87$. At fixed sensitivity this corresponds, as a consequence of the fact that the strain sensitivity goes as $T^{1/4}_{FFT}$ and the computing cost as $T^3_{FFT}$ (considering only the first order spin-down), to a reduction in computing cost of more than a factor of 5.  These results have been obtained by making a comparison with a specific implementation of the standard Hough transform described in \cite{c6c7} where over-resolution is not applied. A comparison with other implementations could give different results. The effect of frequency over-resolution on the digitization loss for each of the parameters is discussed in Sec.\ref{grid}. Finally the {\it adaptivity}, that is the use of weights to take into account noise non-stationarity and the time-varying detector beam pattern functions, introduced for all the existing implementations of the Hough transform \cite{AdaHou},\cite{hough2},\cite{HoughFFdot},
is very simple and ``natural'' to implement in the FH, as each Hough map is done
for a single sky position and then the weighting due to the detector beam-pattern function is done simply by multiplying all the map pixels by the same number. For the adaptive Hough transform the map amplitude (which is no more an integer number) at a given bin can be written as
\begin{equation}
n=\sum_{i=1}^{N}w_in_i
\label{ncountadap}
\end{equation}
where the weights $w_i$, for a given sky location, depend on the average detector response and average detector noise level in the $i$th FFT.

\begin{figure*}
\includegraphics[width=10cm]{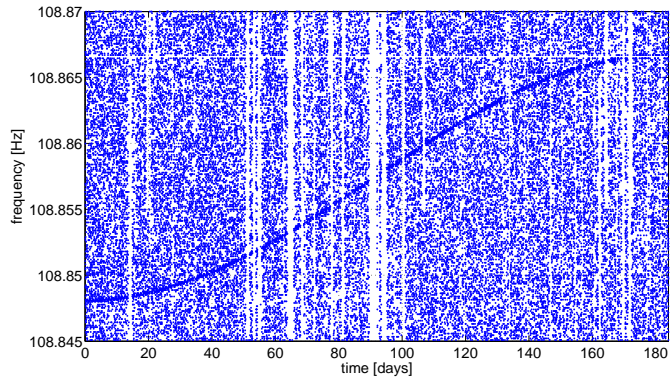} 
\caption{Peakmap around the frequency 
of the HI pulsar3, with $f_0$ = 108.8572 Hz, injected in Virgo VSR2 data. Time is since the beginning of the run. The signal track is clearly visible, due to its very large amplitude.}\label{fig:peakHI3}
\end{figure*}

\begin{figure*}
\includegraphics[width=10cm]{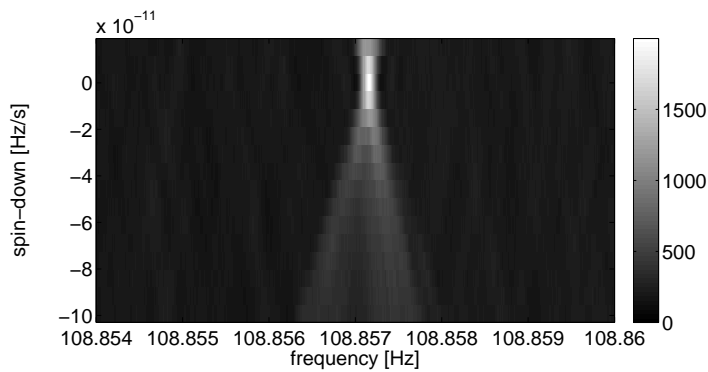} 
\caption{Adaptive FH map around HI pulsar3, at frequency $f_0$ = 108.8572 Hz. The reference time here is the middle of the observation time.}\label{fig:skyHI3}
\end{figure*}

\begin{figure*}
\includegraphics[width=10cm]{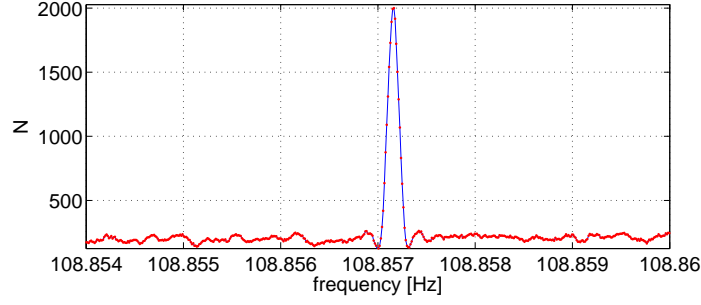} 
\caption{Projection of the Hough map shown in Fig.(\ref{fig:skyHI3}) on the frequency axis.}\label{fig:skyHI3proj}
\end{figure*}

\begin{figure*}
\includegraphics[width=10cm]{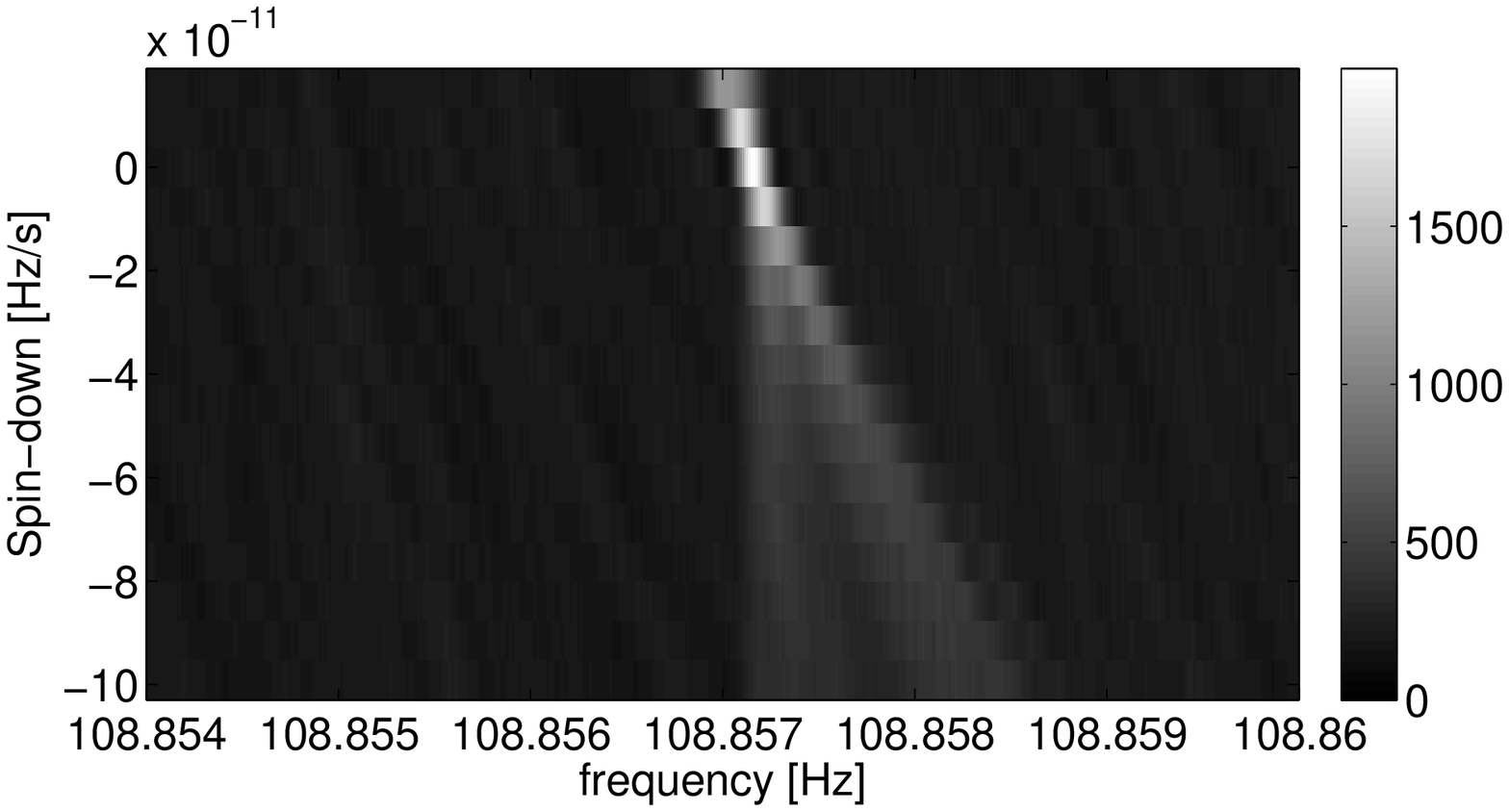} 
\caption{Adaptive Hough map around HI pulsar3, with frequency $f_0$ = 108.8572 Hz, having chosen
the beginning time of the run as reference time.}\label{fig:skyHI3begT}
\end{figure*}

In the following figures we show a few examples of the FH transform, using one of the hardware injections (HI) in Virgo VSR2 run, that is CW injected through the detector hardware for testing purposes.
The so-called pulsar3 has been injected to simulate a CW with
frequency $f_0$ = 108.8572 Hz at epoch MJD 52944, spin-down 
$\dot{f}_0=-1.46 \cdot 10^{-17}$ Hz/s and coming from right ascension $\alpha=178.37^o$ and
declination $\delta=-33.43^o$.
The spin-down is practically zero given the resolution 
we would have in the analysis of VSR2 data.
The amplitude for this signal, $h_0=8.3 \cdot 10^{-24}$, is quite large given
the sensitivity of VSR2 data in that frequency region.
Fig. \ref{fig:peakHI3}  shows the peakmap around the frequency of the HI: the signal track on the time/frequency plane is clearly visible by eye.
Fig. \ref{fig:skyHI3} shows the Adaptive FH map around the HI, using the parameter space grid described in Sec.\ref{grid}. 
The signal parameters are identified by the pixel in the map with the highest number count. Fig. \ref{fig:skyHI3proj} shows the projection
of the Hough map on the frequency axis. The presence of the signal is very well evident also in this plot.
To build the map of Fig. \ref{fig:skyHI3} the reference time has been taken in the middle of VSR2 run. This choice minimizes the uncertainty on source parameters. As an example, in Fig. \ref{fig:skyHI3begT} the Hough map obtained by taking the reference time at the beginning of the run is shown.  

\section{\label{grid}Coarse grid in the parameter space}
In this section we discuss how to build the {\it coarse} grid in the parameter space, frequency, spin-down and sky position, where coarse candidates are selected, as anticipated in Sec. \ref{hier_scheme}.
This is partly based on the results of \cite{HoughFFdot}, with some important improvements. The {\it refined step}, around the candidates found, is described 
in Sec. \ref{refinedgrid}. The use of coarse and refined grids has been also adopted in \cite{ref:eh2013}, but in the context of a different analysis pipeline and with relevant differences in the implementation. Two important points are: a) we use over-resolution in frequency already at the coarse step, without increasing the computational load of the analysis; b) while we build a refined grid only around coarse candidates, in \cite{ref:eh2013} the refined grid covers the whole parameter space. 
\subsection{Grid in frequency}
The ``natural'' grid step in frequency, $\delta f=\frac{1}{T_{FFT}}$, is fixed when constructing the SFDB. 
As shown in \cite{HoughFFdot}, however, the transformation from the peakmap to the Hough plane is not computationally bounded 
by the size of the frequency bin, as it only affects the size of the
Hough map.
As already mentioned in Sec.\ref{freqhough} this means that we can increase the frequency resolution to reduce the digitalization loss. 
To quantify the effect and to do a reasonable choice of the frequency 
over-resolution factor, simulations have been done \cite{HoughFFdot}, by
studying the loss for injected signals in the absence of noise.
The study has lead to identify as a reasonable choice a frequency over-resolution 
factor $K_f$ = 10, both for the coarse and the refined steps. The actual frequency bin width is then
\begin{equation}
\delta f_H = \frac{\delta f}{K_f}
\label{eq:freqres}
\end{equation}
With this choice, in the case of $T_{FFT}=1024$ s, 
the frequency digitalization loss, in amplitude, is of about $3.6\%$, which has to be compared to a loss of $\sim 12\%$ for $K_f=1$. The number of frequency bins in the full band from 0 to $\frac{1}{2\delta t}~Hz$ is
\begin{equation}
N_f=K_f\frac{T_{FFT}}{2\delta t}=\frac{1}{2\delta t\cdot \delta f_H}
\label{nfreqbin}
\end{equation}

\subsection{Grid in spin-down}
\begin{figure*}
\includegraphics[width=8cm]{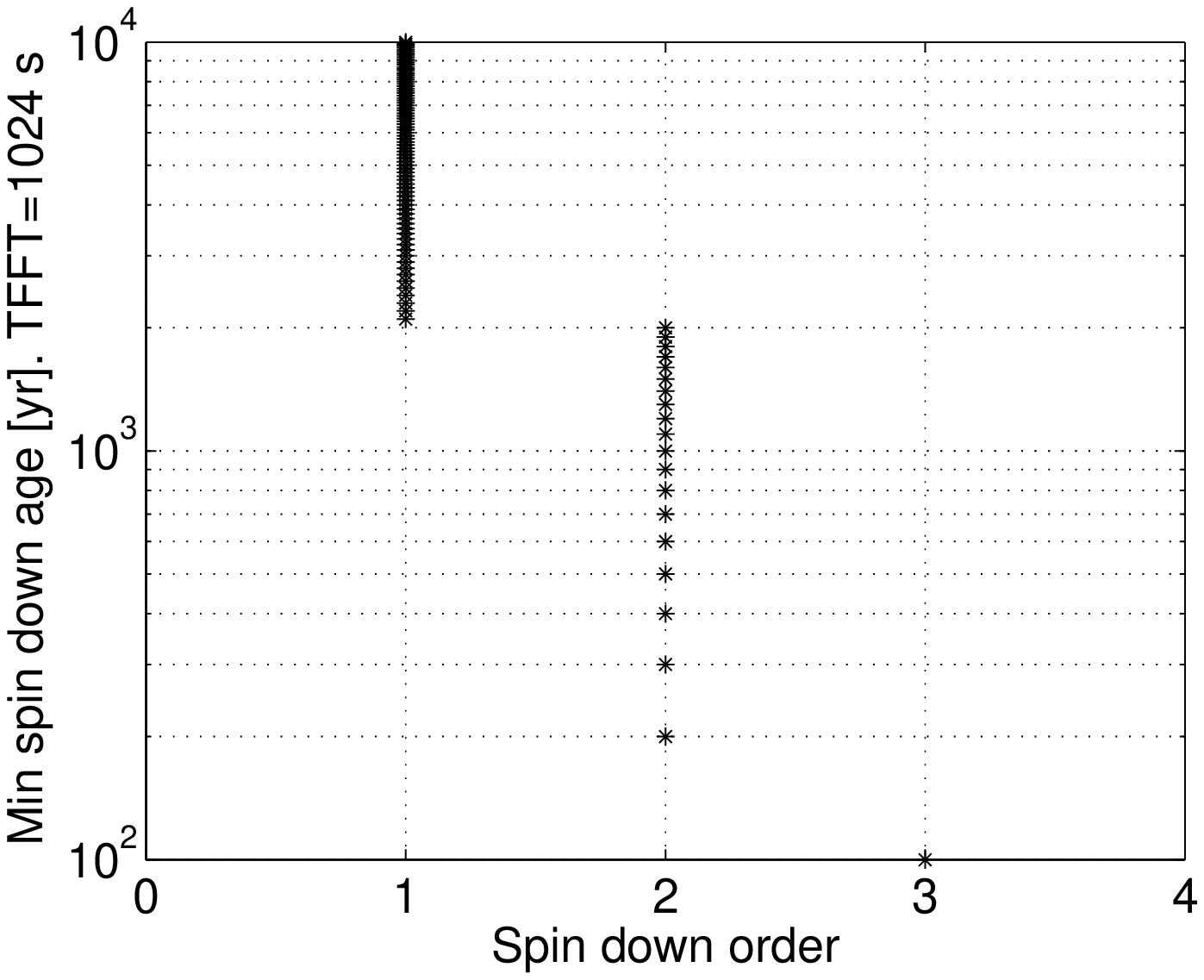}
\includegraphics[width=8cm]{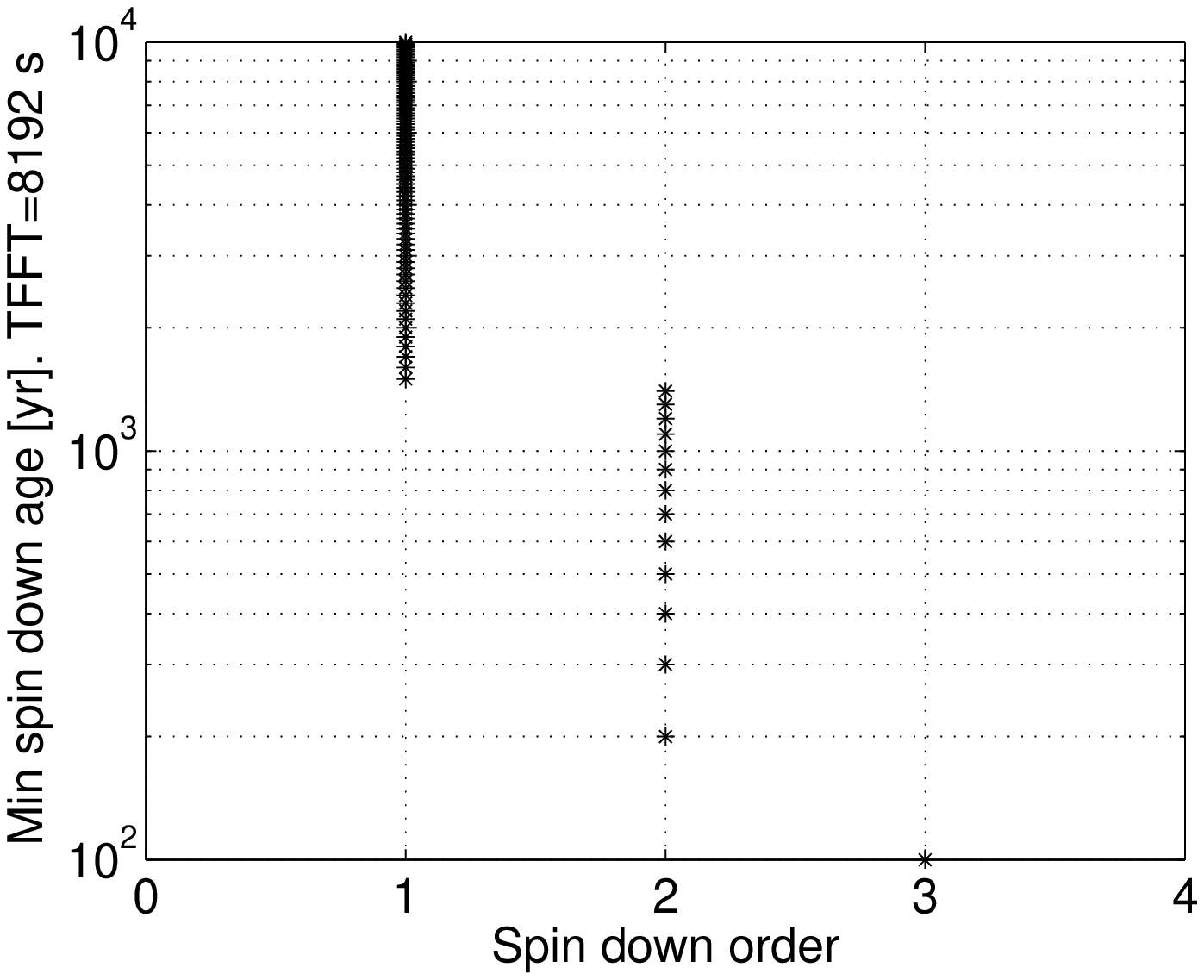}
\caption{Minimum spin-down age $\tau_{min}$ as a function of the spin down order $j$, 
for an observation time of $T_{obs}$= 1 year and two different FFT durations, $T_{FFT}$= 1024 s (maximum frequency 2048 Hz) and 8192 s (maximum frequency 128 Hz).}  
\label{fig:MINSD}
\end{figure*}

The ``natural'' step for spin-down of order $j$, $\delta {f}^{(j)}$, is computed by imposing that the associated frequency variation over the observation time $T_{obs}$ is of one bin:
\begin{equation}
\frac{\delta {f}^{(j)}}{j!}T^j_{obs}=\delta f
\label{sdstep}
\end{equation}
that is $\delta \dot{f}\equiv \delta {f}^{(1)}=\frac{\delta f}{T_{obs}}$, $\delta \ddot{f}\equiv \delta {f}^{(2)}=2\frac{\delta f}{T^2_{obs}}$ and so on.  
As reported in \cite{sergioqui}, the number of values of spin-down values of order $j$ can be determined by using the
following equation:
\begin{equation}
N^{(j)}_{sd}= \frac{T_{FFT}}{\delta t} \left(\frac{T_{obs}}{\tau_{min}}\right)^j
\label{eq:Nsd}
\end{equation}
where $\tau_{min}=MIN\left(\frac{f_0}{\dot{f}_0}\right)$ is the minimum spin-down age considered in the analysis. 
 The total number of points in the spin-down space is
\begin{equation}
N_{sd}=\prod_{j\le j_{max}} N_{sd}^{(j)}
\label{eq:Nsdtot}
\end{equation}
where the maximum order $j_{max}$ to be considered is the last one which has $N^{(j)}_{sd} \ge 1$. The choice of $\tau_{min}$ has a relevant impact on the computational load of the analysis.   
Fig. \ref{fig:MINSD} gives, for different values of $\tau_{min}$, between 100 years and 10000 years, the corresponding maximum spin-down order that must be taken into account  
for an observation time $T_{obs}$= 1 year and
two different sets of FFTs, of duration respectively 1024 s (maximum frequency 2048 Hz)
and 8192 s (maximum frequency 128 Hz). 
From the figures we see that, for 1 yr observation time, the minimum spin-down age in order to have only the first spin-down parameter is 2100 yrs for $T_{FFT}$= 1024 s and 1500 yrs for $T_{FFT}$= 8192 s. Current analysis procedures includes only the first order spin-down. 

The first order spin-down resolution can be generalized as follows:
\begin{equation}
\delta \dot{f} = \frac{\delta f}{T_{obs}K_{\dot{f}}}
\end{equation}
allowing for an over-resolution factor $K_{\dot{f}}$.
The choice we have done for the coarse step of the search is $K_{\dot{f}}=1$. 
The use of an over-resolution factor for the spin-down would in fact have a relevant  
impact on the computing load (as the evaluation of the spin-down has to be done 
by cycling on all the values). Besides this, the amplitude 
digitalization loss is $\sim 3.6\%$ for $K_{\dot f}=1$, small enough to justify the choice we have done. $K_{\dot{f}}>1$ will be used in the refined step.

Instead of fixing a value for $\tau_{min}$ and then use Eqs.(\ref{eq:Nsd},\ref{eq:Nsdtot}) to compute the corresponding number of spin-down values, we could fix the number of spin-down value $N_{sd}$ we want to search. The corresponding minimum spin-down age would then be given by
\begin{equation}
\tau_{min}=\frac{2f_{max}}{N_{sd}\delta \dot{f}}
\end{equation}
where $f_{max}$ is the maximum frequency of the search band.
\subsection{\label{subsec:skygrid}Grid in the sky}
The procedure to construct the sky grid is based on what described in \cite{HoughFFdot}. 
Let us consider two hypothetical sources, emitting a signal at the same frequency $f_0$, having the same ecliptic latitude $\beta$ and a small angular separation in the ecliptic longitude, $\gamma$. 
Due to the detector motion, the separation between the two sources can be seen as a time delay $\Delta t\approx \gamma/\Omega_{orb}$, where $\Omega_{orb}$ is the Earth orbital angular velocity (we are neglecting the Earth rotation). The signals they emit are subject to the Doppler effect, described by Eq.(\ref{eq:fdopp}), so that the frequency at the detector is
\begin{equation}
f(t)\simeq f_0 \left(1+\frac{\vec{v}
\cdot \hat{n}}{c}\right) \approx f_0 \left(1+\frac{\Omega_{orb} R_{orb} \cos \beta \sin(\Omega_{orb}t)}{c}\right)
\label{fv}
\end{equation}
where $R_{orb}$ is the radius of the Earth orbit.
The observed frequency variation during $\Delta t$ is given by
\begin{equation}
\frac{df}{dt} \Delta t \approx f_0 \frac{\Omega^2_{orb} R_{orb} \cos \beta \cos(\Omega_{orb}t)}{c} \Delta t
\end{equation}
The maximum value of this variation is:
\begin{equation}
\Delta f_{max}=  f_0 \frac{\Omega_{orb} R_{orb} \gamma \cos \beta}{c}
\end{equation}
If we fix $\Delta f_{max} = \delta f$ 
we find the angular resolution along the longitude which is, in radians:
\begin{equation}
\delta \lambda\equiv \gamma= \frac{c}{f_0 \Omega_{orb} R_{orb} T_{FFT} \cos \beta}= 1/(N_D \, \cos\beta)
\label{gammalong}
\end{equation}
where $N_D$ is 
\begin{equation}
N_D=\frac{f_0 \Omega_{orb} R_{orb} T_{FFT}}{c}
\label{eq:ND}
\end{equation}
We can derive this last equation also considering the maximum Doppler band, $B_D=\frac{f_0 \Omega_{orb} R_{orb}}{c}$, and noticing that
$N_D=\frac{B_D}{\delta f}$ is the number of frequency bins in it.  
We now repeat the same reasoning supposing
the two sources at the same frequency $f_0$ and same ecliptical longitude $\lambda$. The derivative of the frequency with respect to the latitude $\beta$ is
\begin{equation}
\frac{df}{d\beta} =-\frac{f_0 \Omega_{orb} R_{orb} \sin \beta \sin(\Omega_{orb}t)}{c}
\end{equation}
The frequency variation corresponding to a small angular separation $\gamma '$ along the ecliptical declination is $\frac{df}{d\beta} ~\gamma'$,
with maximum value  
\begin{equation}
\Delta f_{max}=|\frac{df}{d\beta} ~\gamma'|_{max} =|\frac{f_0 \Omega_{orb} R_{orb} \sin \beta ~\gamma'}{c}|
\end{equation}
As before, imposing $\Delta f_{max}= \delta f$, we obtain the angular resolution along the declination: 
\begin{equation}
\delta \beta\equiv \gamma'= \frac{c}{f_0 \Omega_{orb} R_{orb} T_{FFT} \sin \beta}= 1/(N_D \, \sin \beta)
\label{gammalat}
\end{equation}
Using Eqs.(\ref{gammalong},\ref{gammalat}) we construct the grid on the sky, see Appendix \ref{A2} for some implementative details of the procedure.
The points of the grid are not uniformly distributed. 
With a simulation we have estimated 
the number of points in the grid $N_{sky}$ which is, in the high
frequency limit,
\begin{equation} 
N_{sky} \simeq 4\pi K_{sky} {N_D^2} 
\label{eq:Nsky}
\end{equation}
$K_{sky}$ is an over resolution factor, 
which can be chosen to be greater than 1, to enhance the efficiency but unfortunately also the number of
artifacts, or less than 1, to save
computing cost and to reduce the number of artifacts, obviously worsening the efficiency.
By ``artifacts'' here we mean the spurious combinations of frequency and spin-down
which produce candidates all due to one single ``true'' signal.
In \cite{HoughFFdot} we have estimated the loss of sensitivity due to the discretization 
of the sky. For a fixed $K_{sky}$ this a function of the frequency over-resolution factor $K_f$. In particular the amplitude loss for $K_{sky}=1$, which is our standard choice for the coarse grid, is about $10\%$ using $K_f=10$ while it would be $\sim 14\%$ when no frequency over-resolution is done as in the standard Hough transform. For $K_{sky}=1$ the number of points given by Eq.(\ref{eq:Nsky}) is a factor of $\pi$ smaller than for the standard Hough transform.   

\begin{figure*}
\includegraphics[width=10cm]{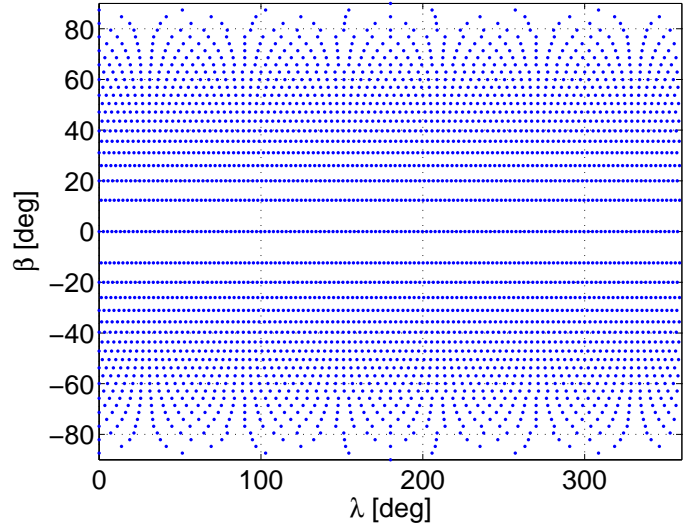}
\caption{Sky grid in ecliptical coordinates, for $T_{FFT}$=1024 s, frequency $f_0=200$ Hz and $K_{sky}=1$. Each point in the plot defines the center of a sky cell.}  
\label{fig:gridsim1}
\end{figure*}

\begin{figure*}
\includegraphics[width=8cm]{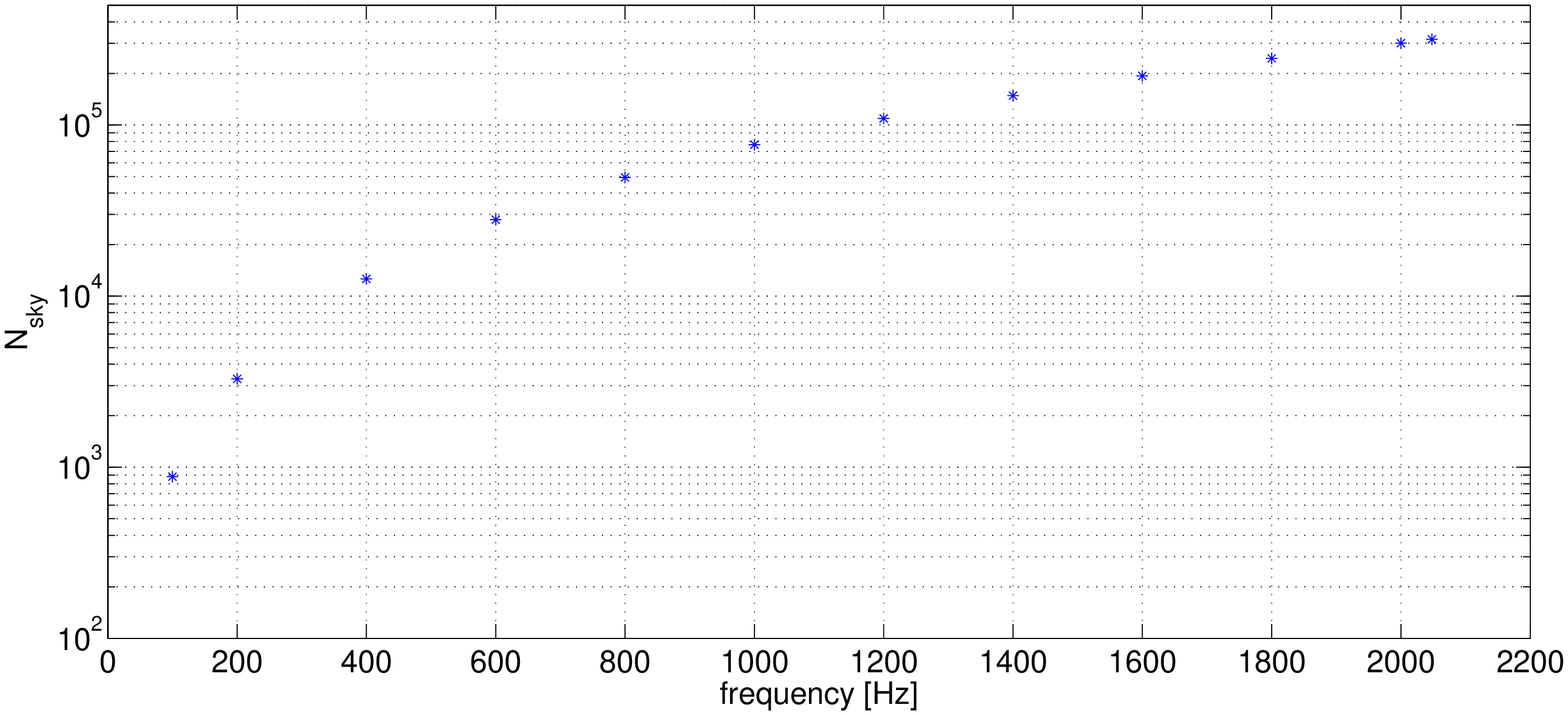}
\includegraphics[width=8cm]{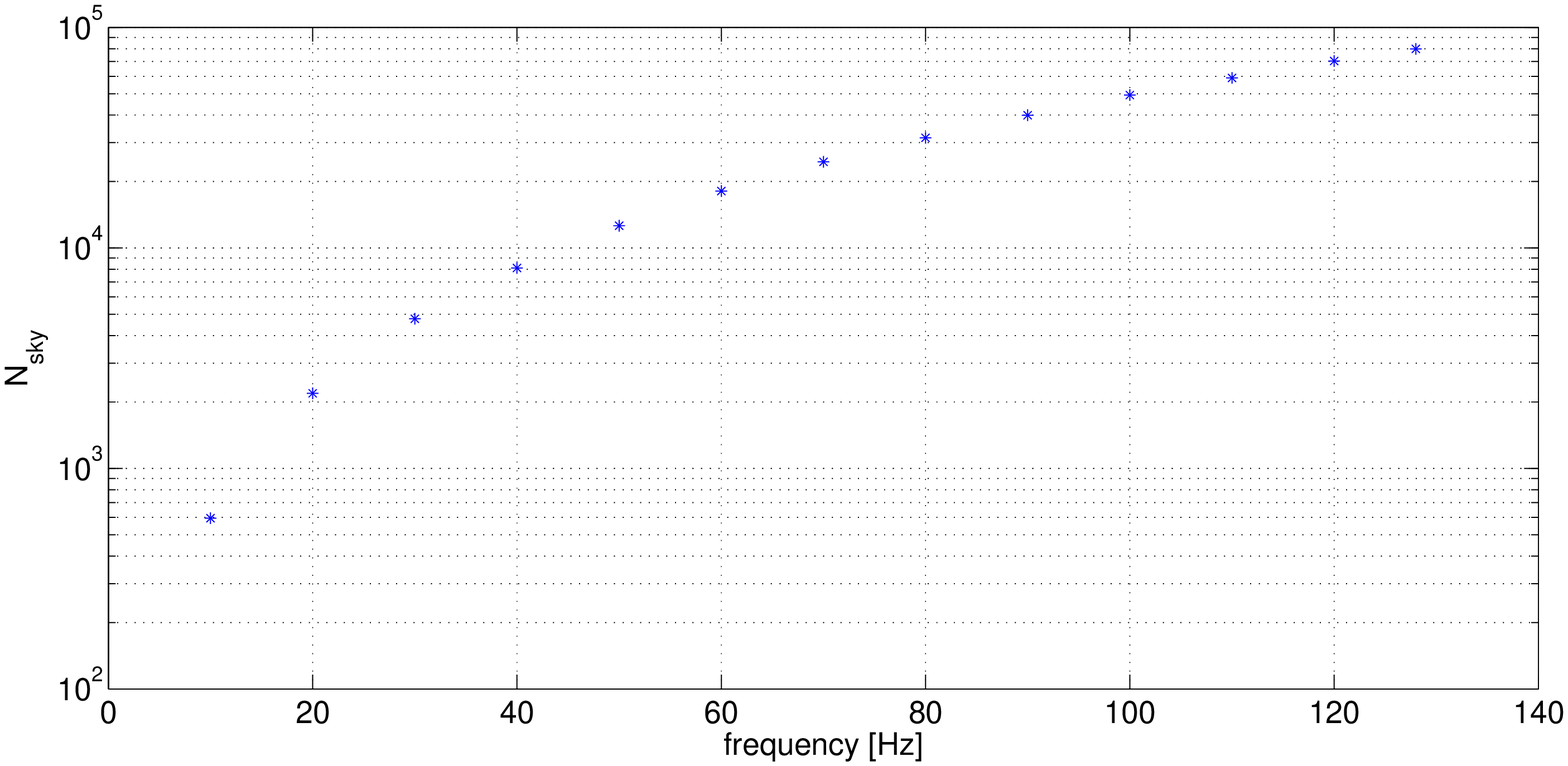} 
\caption{Number of points in the sky grid, $N_{sky}$, constructed with $K_{sky}$=1, as
a function of the frequency in the two cases of $T_{FFT}$ =1024 s (left) 
and 8192 s (right).}
\label{fig:NSKY}
\end{figure*}

Fig.\ref{fig:gridsim1} shows the sky grid for a (maximum) frequency of 200 Hz and $T_{FFT}=1024$ s. 
Fig. \ref{fig:NSKY} shows the number $N_{sky}$
of points in the grid as a function of the frequency, in the two cases of
 $T_{FFT}$ =1024 s and 8192 s.

\section{\label{cand}Selection of first level candidates}

As briefly outlined at the beginning of Sec.\ref{hier_scheme}, after the Hough transform has been computed for a given dataset, a number of first level candidates is selected, taking those with highest significance measured, for instance, by the critical ratio (CR), defined in Sec.\ref{sens}, and which will be used for the next steps of the analysis. This number is chosen as a compromise between the need from one hand to have a manageable amount of candidates and to the other to limit the sensitivity loss that the selection implies, see Sec.\ref{sens}. In fact choosing a reasonable threshold on the CR we can expect that most of the selected candidates are false. In order to reduce the false alarm probability, another set of candidates is selected analyzing a different dataset, belonging to the same detector or not, and coincidences among the two sets of candidates are done. Indeed, given the persistent nature of CW, a signal producing a candidate in a dataset will produce a candidate with (approximately) the same parameters in another dataset, even if this covers a different time span. In principle, that is neglecting the fact that due to the noise the candidates corresponding to a signal could have slightly different parameters in the two analyses and then that a coincidence window must be used, the number of coincidences is given by, see Appendix \ref{sec:theory_coin}   
\begin{equation}
N_{c}\approx \frac{N_{cand1} \cdot N_{cand2}}{N_{tot}}
\label{coinci}
\end{equation}
where $N_{cand1},N_{cand2}$ is the number of candidates selected on the two datasets, while  $N_{tot}=N_f\cdot N_{sky}\cdot N_{sd}$ is the total number of points in the source parameter space, assumed to be the same for the two analyses. By using Eqs.(\ref{nfreqbin},\ref{eq:Nsd},\ref{eq:Nsky}) we can write
\be
N_{tot}\approx 5.6\pi\cdot 10^{-9}K_{f}K_{sky}\left(\frac{T_{FFT}}{\delta t}\right)^{3+j_{max}}\prod_{j\le j_{max}}
\left(\frac{T_{obs}}{\tau_{min}}\right)^j
\label{Ntot}
\ee
where the productory is done over all values of $j\le j_{max}$ such that $N_{sd}^{(j)}$, defined by Eq.(\ref{eq:Nsd}), is $\ge 1$.
For instance, taking $\delta t=1/4096 ~s$, $T_{FFT}=1024~s$, $T_{obs}=1~yr$ and
$\tau_{min}=10^3~yr$ we have $j_{max}=2$ and $N_{tot}\simeq 2.28\cdot 10^{17}$ where we have used $K_f=10$ and $K_{sky}=1$, see also Tab.\ref{thres}. If we decide to select $10^9$ candidates in each dataset we would have, theoretically, about 4 coincidences if the noise was Gaussian. 
In fact, we fix the theoretical number of coincident candidates we want to follow-up, $N_c$, and determine the corresponding number of candidates to be selected in each dataset.
Assuming for simplicity $N_{cand1}=N_{cand2}=N_{cand}$, from Eq.(\ref{coinci}) we get:
\begin{equation}
N_{cand} = \sqrt{N_{c} N_{sky} N_{f} N_{sd}}
\label{eq:ncand}
\end{equation}
In practice, the full frequency range considered in the analysis is split, for computational efficiency reasons, in a number $n_{band}$ of non-overlapping bands (e.g. 1 Hz wide) each of which is analyzed separately and independently of the others. For a given band width, the number of points in the corresponding portion of the parameter space increases with the square of the band maximum frequency, see Eq.(\ref{eq:ND}, \ref{eq:Nsky}). Let us then consider the last band of the full frequency range we are exploring, i.e. that with the highest frequency, and fix the number $N_{c,max}$ of surviving candidates we want to have after coincidences with the corresponding band of another dataset. Let also indicate with $N_{sky,max}$ the number of sky points in this band. The corresponding number of candidates to be selected before coincidences, in order to have $N_{c,max}$ coincident candidates, is given by Eq.(\ref{eq:ncand}) replacing $N_c$ with $N_{c,max}$ and $N_{sky}$ with $N_{sky,max}$:
\begin{equation}
N_{cand,max} = \sqrt{N_{c,max} N_{sky,max} N_{f} N_{sd}}
\label{eq:ncandmax}
\end{equation}
We now impose that the number of coincidences in {\it all} the bands is the same, that is $N_{c;i}=N_{c,max}$, where the index $i=1,...n_{band}$ indicates the $i$th band. Hence, the number of candidates to be selected in the $i$th band is
\begin{equation}
N_{cand;i} = \sqrt{N_{c,max} N_{sky;i} N_{f} N_{sd}}
\label{eq:ncandband}
\end{equation}
where $N_{sky;i}$ is the number of sky cells in the $i$th band and we are assuming for simplicity that all the bands have the same width so that $N_f$ and $N_{sd}$ are constant.
In order to have a uniform number of coincidences in each frequency band and for each band in each sky cell, the number of candidates that will be selected for each cell of the sky is given by
\begin{equation}
N^{(cell)}_{cand;i} = \frac{N_{cand;i}}{N_{sky;i}}=\sqrt{\frac{N_{c,max}N_{f} N_{sd}}{N_{sky;i}} }
\end{equation}
In Sec.\ref{clustandcoin} we will see in some more detail how coincidences are done in practice. 

\begin{figure*}
\includegraphics[width=10cm]{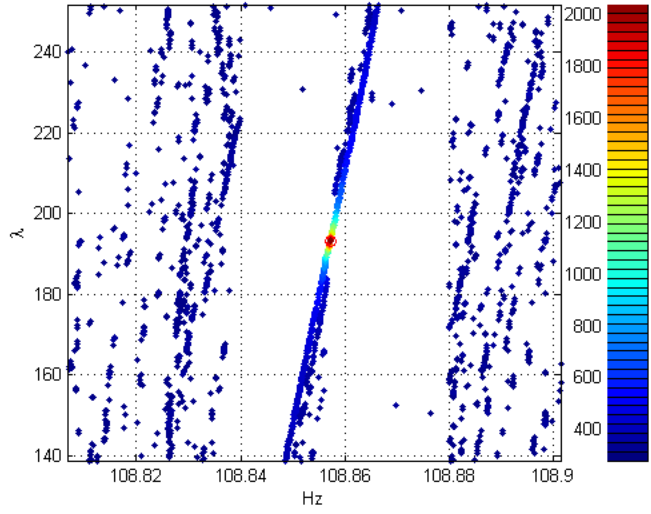}
\caption{The figure shows the Hough map amplitude for the selected candidates, for
a range of values of the ecliptical longitude $\lambda$, around the frequency of the HI pulsar3. The empty regions around the HI are
due to the selection procedure, as explained in the text. 
The few candidates appearing around the HI are of 
``second order''.}   
\label{fig:canSEL}
\end{figure*}

We now focus attention on the practical procedure to select candidates from a Hough map. Once we have fixed the size of the frequency bands on which to run the search, 
the total number of candidates to be selected in each of them $N_{cand,i}$ and
the number of candidates in each cell of the sky, $N^{(cell)}_{cand;i}$, we face the problem
of not being blinded by the presence of disturbances, 
which could still pollute sub-bands of the $i$th band, even after having performed all the cleaning steps described in Sec.\ref{pmclean}. 
We have designed a procedure for candidate selection to this purpose. 
For each sky cell, we divide the $i$th band  
into $n_{sb}=N^{(cell)}_{cand;i}$ sub-bands and select the most significant candidate in each of them, for all the possible spin-down values. In this way the selection of a uniform distribution of candidates is done in each band and the blinding effect due to possible large disturbances is eliminated.
A further step can consist in the selection of ``second order'' candidates. Once 
the highest candidate in each sub-band has been selected, an exclusion region of e.g. $\pm 4$ frequency bins around it is established. We can now look for the second highest candidate in that sub-band 
and select it only if well separated in frequency
from the first one, e.g. by at least $\pm 8$ frequency bins. In this way we expect in general
to select 2 candidates per sub-band and to have 1 candidate only when the 
highest candidate is due to a big disturbance, or a particularly strong HI, as in the following example, which would produce several other neighbouring candidates. This procedure would imply a nearly doubling of the final number of candidates with respect to the initial choice. 
Fig.(\ref{fig:canSEL}) gives and example of what happens around one HI, pulsar3, which is well visible and 
identified by the highest (red) pixel. The figure shows the Hough map number count of the candidates, for a range of values of the ecliptical longitude $\lambda$, around the frequency of the HI. In this example the number of sub-bands is $n_{sb}=23$, each of width 0.043 Hz. 
This explains the presence of (almost) empty regions around the pixels due to the HI.

\section{\label{refinedgrid}Refined Grid in the parameter space}
As already briefly mentioned in Sec. \ref{hier_scheme}, once a candidate is selected using the {\it coarse} grid in the parameter space, the FH transform is run again in a small volume of the parameter space around it using a {\it refined} grid. For each coarse candidate only one refined candidate is selected. The refinement has not any influence on the search sensitivity, which is fixed once the candidates are selected. On the other hand it is very important when coincidences among candidates found in different datasets are done. In fact, it allows to reduce the uncertainty in the candidate parameters and consequently to use a smaller coincidence window, which implies a smaller number of coincident candidates.
The construction of the refined grid is described in the following.
\subsection{Refined grid in frequency}
As already explained in Sec. \ref{grid}, the grid in frequency uses an over-resolution factor, 
which we have fixed to 10, both for the coarse and refined steps. No further refinement is needed. A range of $\pm 1$ coarse bins are considered for the refinement.
\subsection{Refined grid in spin-down}
\begin{figure*}
\includegraphics[width=10cm]{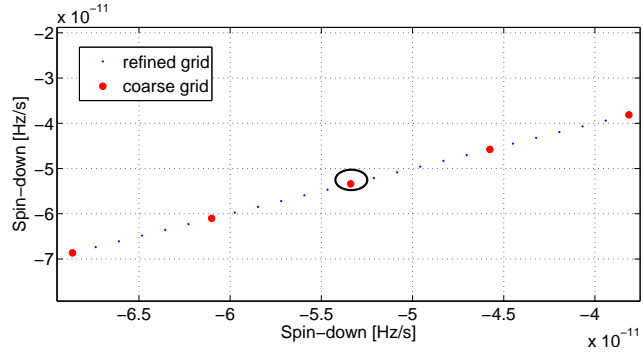}
\caption{Spin-down grid around an hypothetical candidate. Big red dots are points of the coarse grid, small black dots are points of the refined grid. $K_{\dot{f}}\, =6$ in this example.}   
\label{fig:spindownREF}
\end{figure*}

We enhance spin-down resolution by using  $K_{\dot{f}}\, >\, 1$ 
during the refined step.
This is a rather delicate point in view of the coincidence step. The parameters of a candidate refer to a given reference time, typically the middle time of the corresponding dataset. When coincidences among candidates of different datasets are done, the parameters of each pair of candidates must be obviously referred to the same time. In particular, this means that the candidate frequency must be shifted by using the corresponding spin-down value. Then 
the uncertainty in the estimation of the candidate spin-down value, $\delta \dot{f}$, will result in an uncertainty in the estimation of the frequency of the candidate possibly larger than the frequency bin and given by $\Delta f\, =\, \delta \dot{f} \times \Delta T$, being $\Delta T$ the difference between the middle time of a given dataset and the new reference time used for coincidences. It is then clear that the better is the accuracy in spin-down estimation and the better it is, because the resulting uncertainty in frequency will be smaller. The smaller is the uncertainty in frequency and the smaller can be chosen the coincidence window, which will result in a smaller number of coincidence candidates. On the other hand, increasing the spin-down resolution implies a bigger computational load so, as usual, a compromise must be found.  
Fig. \ref{fig:spindownREF} shows, by plotting on both axes the spin-down values,
an example of the coarse grid (red dots) and the refined grid (blue dots),
in the case $K_{\dot{f}}\, =6$, for an hypothetical 
candidate, evidenced by a circle in the plot.
The coarse interval between the spin-down of the candidate and the next value (on both sides) is divided into $K_{\dot{f}}$ pieces. The refined search range includes 
 $2\, K_{\dot{f}}$ on the left of the coarse original value, 
and $(2\, K_{\dot{f}}-1)$ on the right, so that two coarse bins are covered on both sides.
This choice is dictated by the fact that the refinement is in parallel done also on the
position of the source and so a coarse candidate could be found with a refined spin-down value outside the original coarse bin.

\subsection{Refined grid in the sky}
\begin{figure*}
\includegraphics[width=10cm]{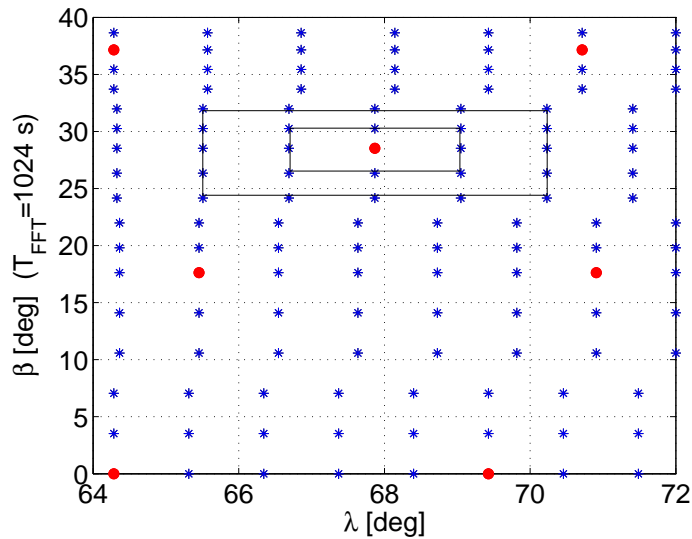}
\caption{An example of refined sky grid. Red dots define points of the coarse grid, black asterisks are points of the refined grid. The black rectangles defines the two ``layers'' that identify the refinement region around an hypothetical candidate.}  
\label{fig:Sky_CoaRefA}
\end{figure*}

The refinement of the sky position of each candidate is done by using a 
rectangular region centered at the candidate coordinates. 
The over-resolution factor, $\hat{K}_{sky}$, is different from  the over-resolution 
$K_{sky}$ in Sect. \ref{grid} as it is a refinement constrained to be symmetric around the candidate. 
The distance between the estimated latitude (longitude) and the next 
latitude (longitude) point in the coarse grid is divided into  $\hat{K}_{sky}$ points, as
shown in Fig.\ref{fig:Sky_CoaRefA}. Here the coarse grid is
indicated by red points and the refined by black asterisks and $\hat{K}_{sky}=5$ in this example which refers to a (maximum) frequency of 100 Hz and $T_{FFT}=1024$ s. For a given coarse candidate the refined coordinates we consider are those forming a number of
``layers'' $N_{layers}$ around the candidate and centered on it, where
$\hat{K}_{sky}=2 \, N_{layers}+1$. $N_{layers}=2$ in the example given.
Fig.\ref{fig:Sky_CoaRefA} shows also the layers around 
an hypothetical candidate and the  $\hat{K}_{sky} \times \hat{K}_{sky}$ 
(25 in this case) refined points in the grid, which are those touched by the black rectangles.

\section{\label{clustandcoin} Candidate clustering and coincidences}
As already explained, coincidences among two, or more, candidate sets are done in order to strongly reduce the false alarm probability. This is a fundamental step in a wide-parameter search to make the next steps of the analysis feasible. 

In fact candidates in each set are organized in {\it clusters}. To define a cluster we first introduce a norm in the candidate parameter space. Given two candidates, each defined by a set of 4 parameter values, $\vec{c}_1=(\lambda_1,~\beta_1,~f_1,~\dot{f}_1)$ and $\vec{c}_2=(\lambda_2,~\beta_2,~f_2,~\dot{f}_2)$ respectively, we define their distance as
\begin{equation}
d=\|\vec{c}_1-\vec{c}_2\|=\sqrt{k^2_{\lambda}+k^2_{\beta}+k^2_f+k^2_{\dot f}}
\label{norm}
\end{equation}
where $k_{\lambda}=\frac{|\lambda_2-\lambda_1|}{\delta \lambda}$ is the difference in number of bins between the ecliptical longitudes of the two candidates, being $\delta \lambda=\frac`{d\lambda_1+d\lambda_2}{2}$ is the mean value of the width of the coarse bins in the ecliptical longitude for the two candidates (which can be different because the resolution in longitude depends on the longitude itself), and similarly for the other terms. A cluster is defined as the subset of candidates such that each of them has a distance from at least another candidate of the same subset less or equal than a given values, e.g. $d\le 4$.   
Clusterization is useful as it may give hints on the common origin of the candidates belonging to the same cluster. For instance a very large cluster or a cluster which candidates have position near the poles is likely due to some disturbance. 

Altough in Sec.\ref{cand} the choice of the number of candidates has been discussed without considering any uncertainty in candidate parameters, in fact when
coincidences are done it is necessary to choose a coincidence window associated to each candidate. Its width is chosen as a compromise between the need to not increasing too much the number of coincident candidates and the need to not discard real signal candidates that, due to noise, could be found with slightly different parameters in the analyzed datasets. In practice, coincidence windows of a few bins for each parameter are a reasonable choice. The number of expected coincident candidates as a function of the coincidence window is given by Eq.(\ref{eq:ncoin_approx}), and will be larger than the number estimated from Eq.(\ref{coinci}). Moreover, if the selection of ``second order'' candidates is done, the actual number of candidates is nearly doubled, see discussion at the end of Sec.\ref{cand}, with a further increase in the number of coincidences. On the other hand, however, in order to largely reduce the number of coincidences due to noise, a possible way to proceed is that of making coincidences among clusters (two clusters are coincident if at least a pair of candidates are coincident) and then considering not all the coincident pairs but only those (one or a few) which are nearest. This clearly implies a reduction of sensitivity. The actual choice of the procedure to be used depends on the characteristics of the data being analyzed.

\section{\label{verifollow} Candidates verification and follow-up}
Surviving candidates after coincidences are subject to a {\it verification} step that allows to furtherly increase confidence in detection or to discard them. 
The verification consists in the application of various criteria not directly to the coincidences but, rather, to the candidates that originated them or even to the peaks in the peakmap that originated the candidates. Among the most important there is a comparison between the signal amplitudes associated to the candidates which generated a given coincidence. If two coincident candidates are due to a real signal we expect the signal amplitude to be the same in the two datasets. The application of this criterium requires a good calibration of the FH transform, that is the knowledge of the relation between the Hough map amplitude and signal amplitude. Another criterium consists in looking at the peaks in the peakmap which originated the coincident candidates. If they are due to a real signal we expect the peaks to be properly distributed in time. For instance, if the peaks which generate a candidate are strongly concentrated in a short period of time this is a clue of its noise origin. One more verification step is based on the detector radiation pattern corresponding to the coincident candidates position. We expect that the number of peaks in the peakmap which contribute to these candidates follows the radiation pattern, with a smaller number of peaks when the detector orientation is ``bad'' and larger when it is ``good''. See also \cite{ref:keit} for another possible candidate verification criterium. 

Candidates which pass also the verification step are subject to a 
{\it follow-up} analysis in which a small portion of the parameter space around each of them is analyzed with a longer coherence time. This implies the construction of a new set of longer FFTs and, possibly, a new Hough transform. See e.g. \cite{ref:loose}, \cite{ref:shal} for proposed follow-up procedures. Details of the follow-up procedure will be discussed elsewhere. It is important to stress that both the verification and the follow-up do not increase the search sensitivity, which is basically set by the initial length of FFTs and by the thresholds used to select peaks in the peakmap and candidates on the Hough map. If a signal is missed at the first Hough step it will be no more recovered. On the contrary, verification and follow-up allow to strongly increase the detection confidence. In particular, increasing the coherence time from, say, $T_{FFT}$ to $T'_{FFT}$ determines a signal-to-noise ratio increase of a factor $\sqrt{\frac{T'_{FFT}}{T_{FFT}}}$ at the follow-up coherent step. Obviously, also a much better determination of the signal parameters is possible.

\section{\label{sens}Search sensitivity estimation}
In this section we compute the theoretical sensitivity of the analysis method, showing in particular the dependency on the thresholds for peaks and candidate selection. Also Receiver Operating Characteristic curves are computed assuming Gaussian noise. Sensitivity loss due to digitizations, discussed in Sec.\ref{freqhough}, is not taken into account here.
\subsection{Threshold for peaks selection}
\begin{figure*}
\includegraphics[width=10cm]{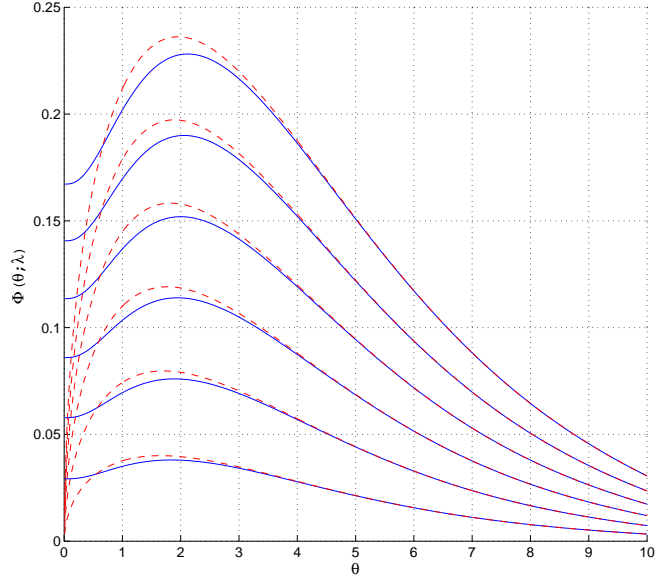} 
\caption{$\Phi$ function, Eq.(\ref{phi}), as a function of the threshold $\theta$ for peak selection
for different values of signal amplitude, from $\lambda=0.1$ (lower curves) to
$\lambda=0.6$ (upper curve), with step of 0.1. Blue continuous curves correspond to the choice of local maxima above the threshold,
while red dashed curves corresponds to the case in which selected peaks are not necessarily local maxima.}\label{fig:phi}
\end{figure*}

Let us introduce the {\it critical ratio}, which is a random variable measuring the
statistical significance of the number count $n$ found in a given pixel of an Hough
map, with respect to the expected value in presence of noise alone:
\be
CR=\frac{n-Np_0}{\sqrt{Np_0(1-p_0)}}
\label{cr}
\ee
where $N$ is the number of FFTs.
The probability $p_0$ depends on the threshold for peak selection, $\theta_{thr}$, see Eq.(\ref{eq:p0}).
The choice of $\theta_{thr}$ influences the search sensitivity and its computational weight. A criterion that can be used for the
choice of the threshold is the maximization of the expectation value of the critical ratio which, assuming a signal of spectral amplitude $\lambda$ is present, is given by:
\be
\mu_{CR}(\theta;\lambda)=
\frac{N(p_{\lambda}-p_0)}{\sqrt{Np_0(1-p_0)}}=\sqrt{N}\Phi(\theta;\lambda)
\label{phi}
\ee
By plotting $\Phi$ as a function of $\theta$ for different values of $\lambda$ we
can decide where to put the threshold, see Fig.(\ref{fig:phi}). 
In principle, the optimal value of the threshold is that maximizing the function
$\Phi$. In practice, given that it is a rather smooth function of $\theta$ we can choose
a value of the threshold slightly larger than that corresponding to the maximum. A reasonable choice is $\theta_{thr}=2.5$, independently of the signal amplitude over a large and reasonable range. This implies a small sensitivity loss (of $\sim
1\%$) and a significant reduction in the expected number of peaks (about a factor of
$2$) with respect to the optimal threshold.  The corresponding probability of selecting a
noise peak is $p_0=P(\theta_{thr};0)=0.0755$. In Fig.(\ref{fig:phi}) the $\Phi$ function is plotted also for the case in which the peaks are selected according to the simpler criteria 
of being above the threshold (not necessarily local maxima), as used in \cite{hough1}. At fixed $\lambda$ the value of $\Phi$ is slightly larger than in the case the local maxima criterium is used meaning a small gain in sensitivity, less than 5$\%$ over a wide and reasonable range of $\lambda$. On the other hand, selecting local maxima, as we do, has two important advantages. First, the number of selected peaks is $\frac{p_0}{e^{-\theta_{thr}}}$ smaller. For instance, for $\theta_{thr}=2.5$ we have a reduction of about 9$\%$. This implies a reduction of the analysis computational load. Second, our criteria is more robust against disturbances. It is quite likely that a disturbance in the data does not affect just a single frequency bin but also its neighbours. Selecting only local maxima clearly makes this problem less relevant.  

\subsection{Threshold for selection of candidates}
\begin{figure*}
\includegraphics[width=10cm]{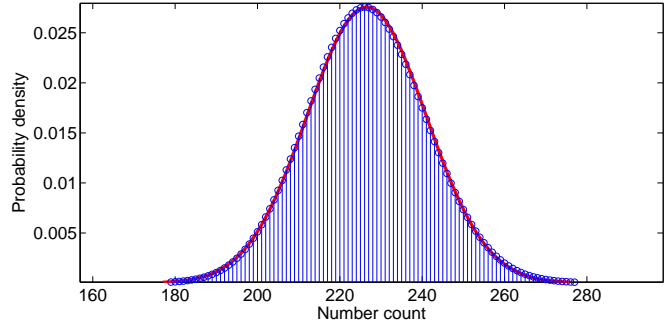} 
\caption{Binomial distribution for $N=3000$ and $p_0=0.075$ (histogram with blue dots) and its
gaussian approximation (continuous line).}\label{bino_gauss}
\end{figure*}

\begin{figure*}
\includegraphics[width=10cm]{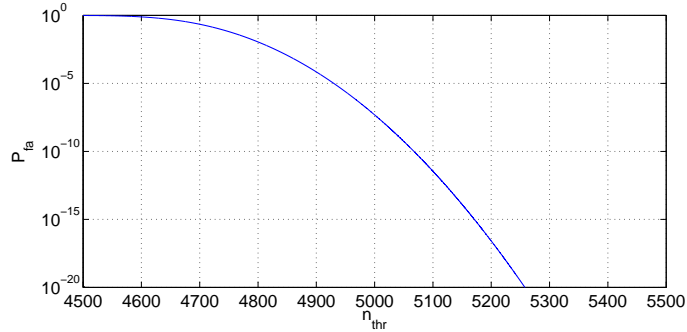} 
\caption{False alarm probability as a function of the
threshold for candidates selection, $n_{thr}$, considering a total observation time of 1 year, $T_{FFT}=1024$ s ($N=61,594$) and $p_0=0.0755$, see Eq.(\ref{pfa_nthr}). With these choices the 
mean value and standard deviation of the number count distribution are respectively, $4650.3$ and $65.6$.}\label{fig:Pfa_N61594}
\end{figure*}

Given the large parameter space we want to explore we need to select a manageable
number of candidates to which further steps of the analysis will be
applied, see Sec.\ref{cand}. For simplicity, in this section we do not take into account that the number of selected candidates will be frequency dependent and that the FFT duration, $T_{FFT}$, will be different in different frequency band. As a result, the sensitivity formula, given by Eq.(\ref{h0min}), depends on the frequency only through the detector noise spectrum $S_n(f)$. In practice, both the threshold on the critical ratio, $CR_{thr}$, and $T_{FFT}$ will be a function of the frequency. 
Let us indicate with $n_{thr}$ the threshold on the number count used to
select candidates on a Hough map. The corresponding false alarm probability is
\be
P_{fa}=\sum_{n=n_{thr}}^N P_n(\theta_{thr};0)
\label{pfa}
\ee
while the false dismissal probability is
\be
P_{fd}=\sum_{n=0}^{n_{thr}-1} P_n(\theta_{thr};\lambda)
\label{pfd}
\ee
where $P_n$ is given by Eq.(\ref{eq:phmsig}).
To write some useful equations we use the Gaussian approximation to the binomial
distribution:
\be
P_n(\theta_{thr};\lambda)=\frac{1}{\sqrt{2\pi
\sigma^2}}e^{-\frac{(n-\mu)^2}{2\sigma^2}}
\ee
where $\mu$ and $\sigma$ are given by Eq.(\ref{houghmean}).
It works fine as long as $N$ is large and $\eta$ is not too near 0 or 1. With our
typical values ($N\approx$ a few thousands, $\eta \ge 0.0755$) the approximation is
very good, see Fig.(\ref{bino_gauss}).
Using this approximation we can compute analytically the threshold on the number of
candidates corresponding to a fixed value of false alarm probability, by writing
Eq.(\ref{pfa}) as:
\be
\int_{n_{thr}}^{\infty} P_n(\theta_{thr};0)dn=P_{fa}
\ee
hence
\be
n_{thr}(N,\theta_{thr},P_{fa})=Np_0+\sqrt{2Np_0(1-p_0)}{\erfc}^{-1}(2P_{fa})
\label{nthr_gauss}
\ee
where ${\erfc}^{-1}$ is the inverse of the complementary error function, which is
defined as $\erfc(x)=\frac{2}{\sqrt{\pi}}\int_x^{+\infty} e^{-t^2}dt$.
Inverting Eq.(\ref{nthr_gauss}) we can write the false alarm probability as a
function of the threshold:
\be
P_{fa}=\frac{1}{2}\erfc{\left(\frac{n_{thr}-Np_0}{\sqrt{2Np_0(1-p_0)}}\right)}
\label{pfa_nthr}
\ee
In Fig.(\ref{fig:Pfa_N61594}) the false alarm probability is plotted
as a function of $n_{thr}$.
The corresponding false dismissal probability, defined by Eq.(\ref{pfd}), is
\be
P_{fd}=\frac{1}{2}{\erfc}\left(\frac{Np_{\lambda}-n_{thr}}{\sqrt{2Np_{\lambda}(1-p_{\lambda})}}\right)
\label{pfd2}
\ee
The detection probability is given by $P_d=1-P_{fd}$. It is easy to see that, as
expected, $P_d=P_{fa}$ for $\lambda=0$.

\begin{figure*}
\includegraphics[width=10cm]{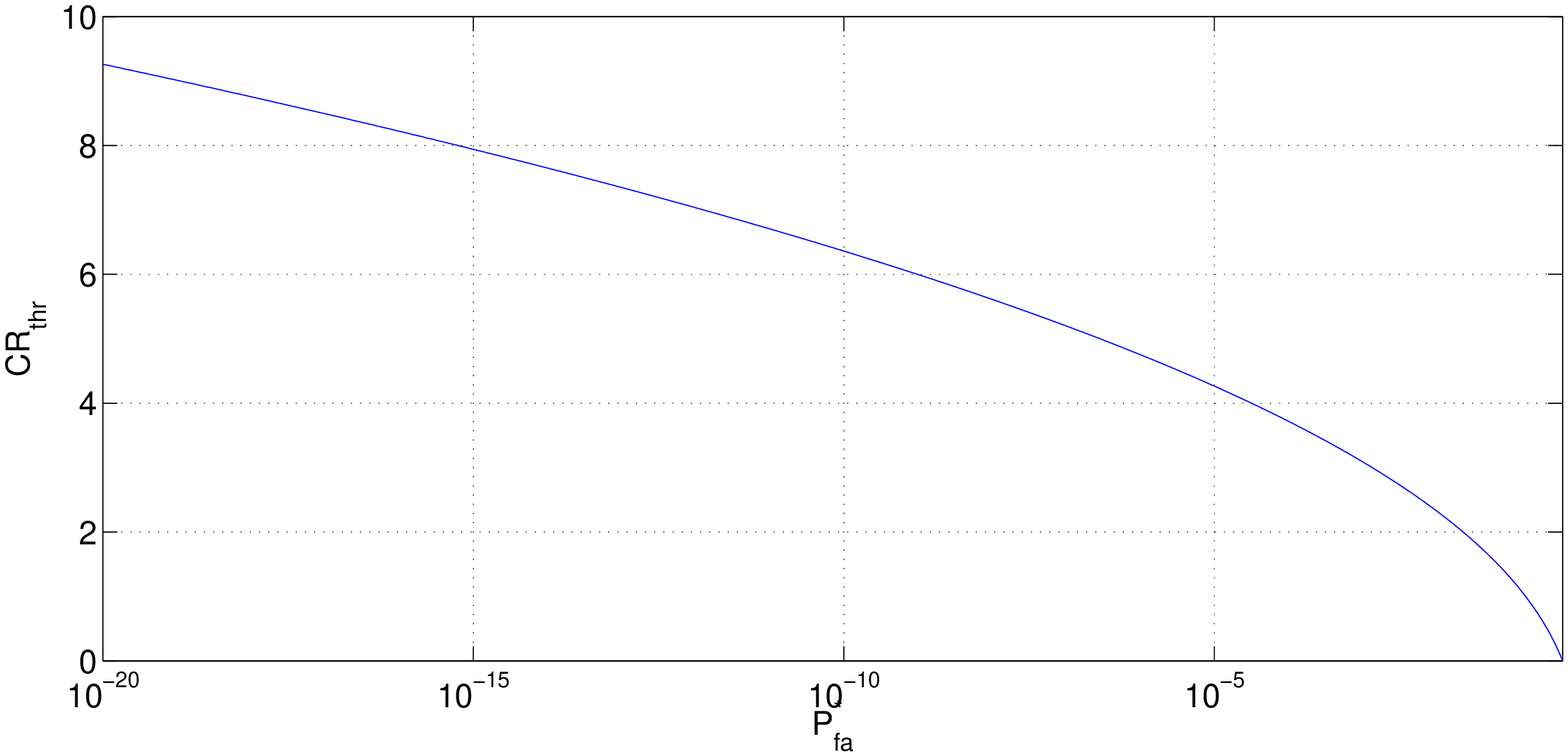} 
\caption{Threshold on the critical ratio for candidate selection as a function of the false alarm
probability, Eq.(\ref{crthr}).}\label{fig:crthr_vs_pfa}
\end{figure*}

\begin{figure*}
\includegraphics[width=10cm]{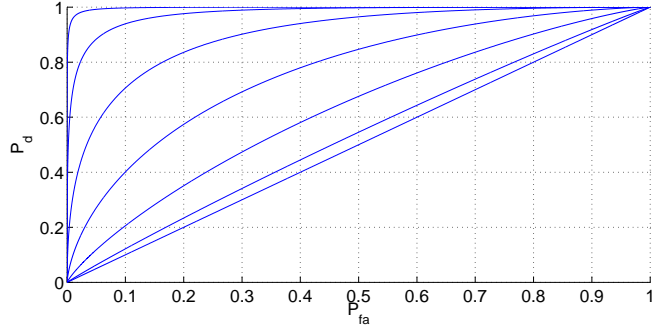} 
\caption{ROC curves for a total observation time $T_{obs}=1~yr$, a noise spectral density $S_n=3.6\cdot 10^{-45}\frac{1}{{Hz}}$, an FFT duration $T_{FFT}=1024$ s ($N=61594$) and $\theta_{thr}=2.5$. From bottom to top, the signal spectral amplitudes
are 0, 0.0014, .0056, 0.0126, 0.0225, 0.0352, 0.0507 which corresponds, through Eq.(\ref{lambda_ave}), to strain amplitudes from 0 to $1.2\cdot 10^{-24}$ with steps of $2\cdot 10^{-25}$.}\label{fig:roc1024}
\end{figure*}

\begin{figure*}
\includegraphics[width=10cm]{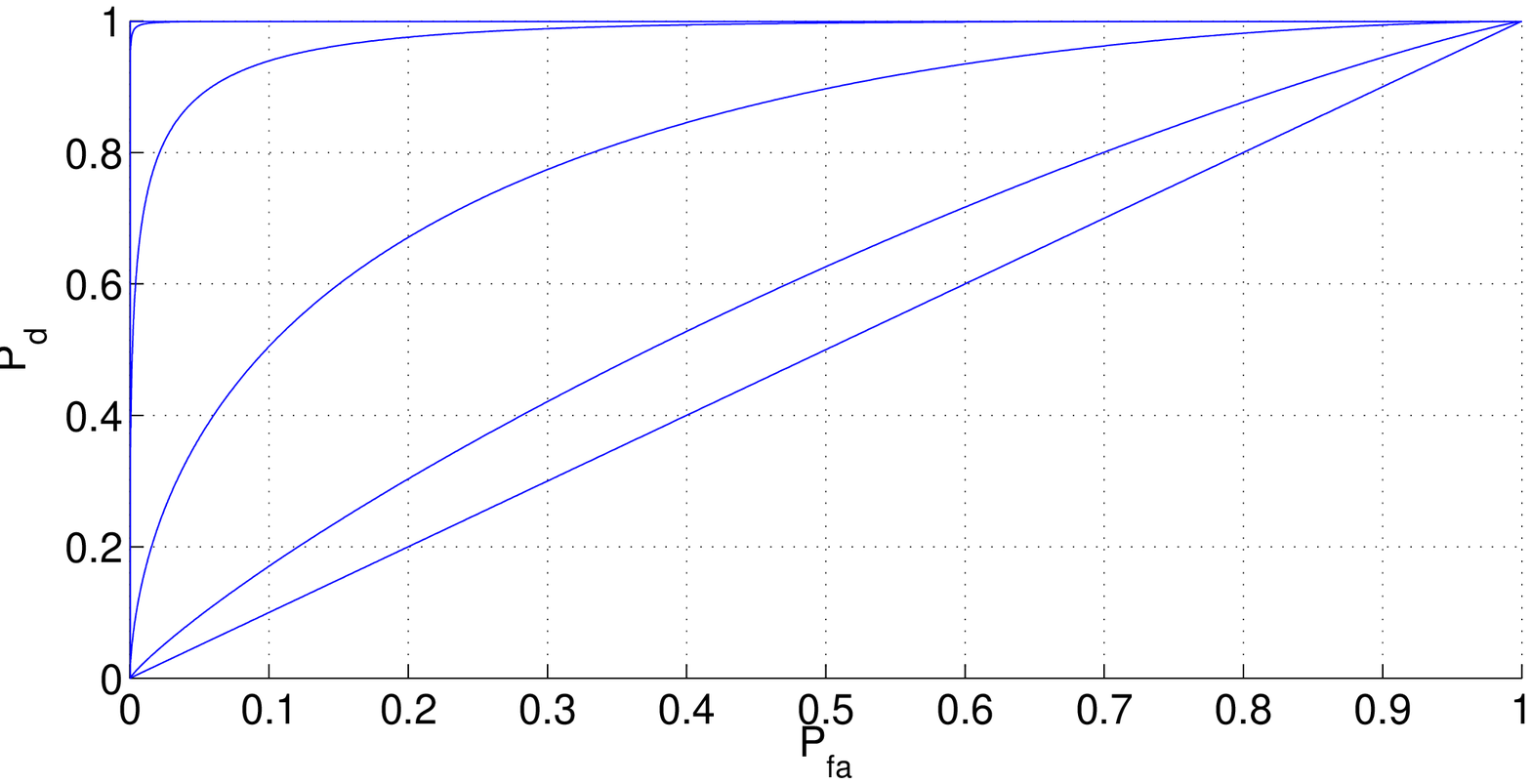} 
\caption{ROC curves for a total observation time $T_{obs}=1~yr$, a noise level $S_n=3.6\cdot 10^{-45}\frac{1}{Hz}$, an FFT duration $T_{FFT}=8192$ s ($N=7699$)and $\theta_{thr}=2.5$. From bottom to top, the signal spectral amplitudes
are 0, 0.0113, .0451, 0.1014, 0.1803, 0.2817, 0.4057 which corresponds, through Eq.(\ref{lambda_ave}), to strain amplitudes from 0 to $1.2\cdot 10^{-24}$ with steps of $2\cdot 10^{-25}$. Note that only the first four curves are visible. For larger signal amplitudes the detection probability is basically one for all false alarm probabilities.}\label{fig:roc8192}
\end{figure*}

\begin{figure*}
\includegraphics[width=10cm]{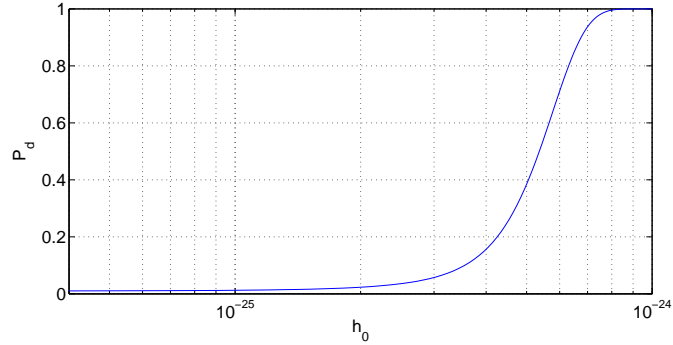} 
\caption{Detection probability as a function of the signal amplitude for false alarm probability $P_{fa}=0.01$ again assuming $S_n=3.6\cdot 10^{-45}\frac{1}{{Hz}}$, an FFT duration of $T_{FFT}=8192$ and $T_{obs}=1~yr$.}\label{fig:pd_vs_h0}
\end{figure*}

In practice, the selection of candidates could be done by putting a threshold on the CR.
We can compute such threshold in the following way. Given the number of
candidates we decide to select, $N_{cand}$, from Eq.(\ref{pfa_nthr}), using the defintion of crtical ratio, Eq.(\ref{cr}), the false alarm probability can be expressed as
\be
P_{fa}=\frac{1}{2}\erfc{\left(\frac{CR_{thr}}{\sqrt{2}}\right)}
\label{fap_cr}
\ee
where $CR_{thr}$ is the value of critical ratio corresponding to $n_{thr}$.
On the other hand, given that the candidates we select are those with the highest CR, it immediately follows that 
\be
P_{fa}=\frac{N_{cand}}{N_{tot}}
\label{pfa_cand}
\ee
where $N_{tot}$ is the total number of points in the source parameter space, given by Eq.(\ref{Ntot}). Note that $N_{tot}$ is computed referring to the coarse grid in the parameter space. As explained in Sec. \ref{refinedgrid}, the refined step is only meant to improve accuracy in candidate parameters and does not affect the search sensitivity.
Using Eq.(\ref{nthr_gauss}), we derive the threshold on the number count and,
with Eq.(\ref{cr}), the threshold on the critical ratio: 
\be
CR_{thr}=\sqrt{2}\erfc^{-1}(2\frac{N_{cand}}{N_{tot}})
\label{crthr}
\ee
In Fig.(\ref{fig:crthr_vs_pfa}) the threshold on the critical ratio is plotted as a
function of the false alarm probability. Note that $CR_{thr}=0$ for $P_{fa}=0.5$.
In Tab. (\ref{thres}) various quantities discussed in this and in the next section are given for different choices of the search parameters. 
\begin{widetext}
\begin{table*}
\caption{\label{thres}Relevant quantities and thresholds for various choices of $T_{FFT}$ and $\tau_{min}$ and assuming $T_{obs}=1~yr$, $K_f=10$, $K_{sky}=1$. For $T_{FFT}=1024 ~s$ the number of (interlaced) FFTs
in $T_{obs}$ is $N=61594$ and we assume to analyze the whole frequency band, between 10 Hz and 2048 Hz, and to select $10^9$ candidates.  For $T_{FFT}=8192~s$ the number of (interlaced) FFTs is $N=7699$ and we assume to analyze the frequency band between 10 Hz and 128 ~Hz, and to select $10^7$ candidates. $\delta t$ is the sampling time, $j_{max}$ is the maximum spin-down order to be considered, $N_{tot}$ is the total number of points in the source parameter space (Eq.(\ref{Ntot})), $P_{fa}$ is the false alarm probability (Eq.(\ref{pfa_cand})), $n_{thr}$ is the corresponding threshold on the Hough map number count (Eq.(\ref{nthr_gauss})) and $CR_{thr}$ is the threshold on the critical ratio used to select candidates (Eq.(\ref{crthr}). $\Lambda_1$ is the sensitivity coefficient appearing in Eq.(\ref{lambdamin}) while $\Lambda$ is the coefficient in Eq.(\ref{h0min}).}
\begin{tabular}{cccccccccc}
$\delta t~[s]$ & $T_{FFT}~[s]$ & $\tau_{min}$ [yr] & $k_{max}$ & $N_{tot}$ & $P_{fa}$ & $n_{thr}$ & $CR_{thr}$ & $\Lambda_1$ & $\Lambda$ \\ \hline \hline
$2.44\cdot 10^{-4}$ & $1024$ & $1,000$ & $2$ & $2.0\cdot 10^{17}$ & $5.00\cdot 10^{-9}$ & $4916$ & $5.73$ & $22.54$ & $13.49$ \\ \hline
$2.44\cdot 10^{-4}$ & $1024$ & $5,000$ & $1$ & $9.7\cdot 10^{15}$ & $1.03\cdot 10^{-7}$ & $4891$ & $5.19$ & $20.89$ & $12.99$ \\ \hline
$3.9\cdot 10^{-3}$ & $8192$ & $1000$ & $2$ & $6.4\cdot 10^{15}$ & $1.56\cdot 10^{-9}$ & $678$ & $5.92$ & $23.12$ & $13.67$ 
 \\ \hline
$3.9\cdot 10^{-3}$ & $8192$ & $5,000$ & $1$ & $6.1\cdot 10^{14}$ & $1.64\cdot 10^{-8}$ & $672$ & $5.53$ & $21.93$ & $13.31$ \\ \hline
\end{tabular}
\end{table*}
\end{widetext}
A useful figure of merit characterizing the performance of a given filtering
procedure is the ROC (Receiver Operating Characteristics), which is a plot, for
different signal amplitudes, of the detection probability $P_d$ as a function of the
false alarm probability $P_{fa}$. In Figs.(\ref{fig:roc1024},\ref{fig:roc8192}) some ROC curves are shown
assuming to have 1 year of data with a noise spectral density $S_n=3.6\cdot 10^{-45}\frac{1}{{Hz}}$ and, respectively, $T_{FFT}=1024$ ($N=61594$) and $T_{FFT}=8192$ ($N=7699$). 
In Fig.(\ref{fig:pd_vs_h0}) the detection probability is plotted as a function of the signal amplitude again assuming $S_n=3.6\cdot 10^{-45}\frac{1}{{Hz}}$, an FFT duration of $T_{FFT}=8192$ and $T_{obs}=1~yr$.

\subsection{Sensitivity}
\begin{figure*}
\includegraphics[width=10cm]{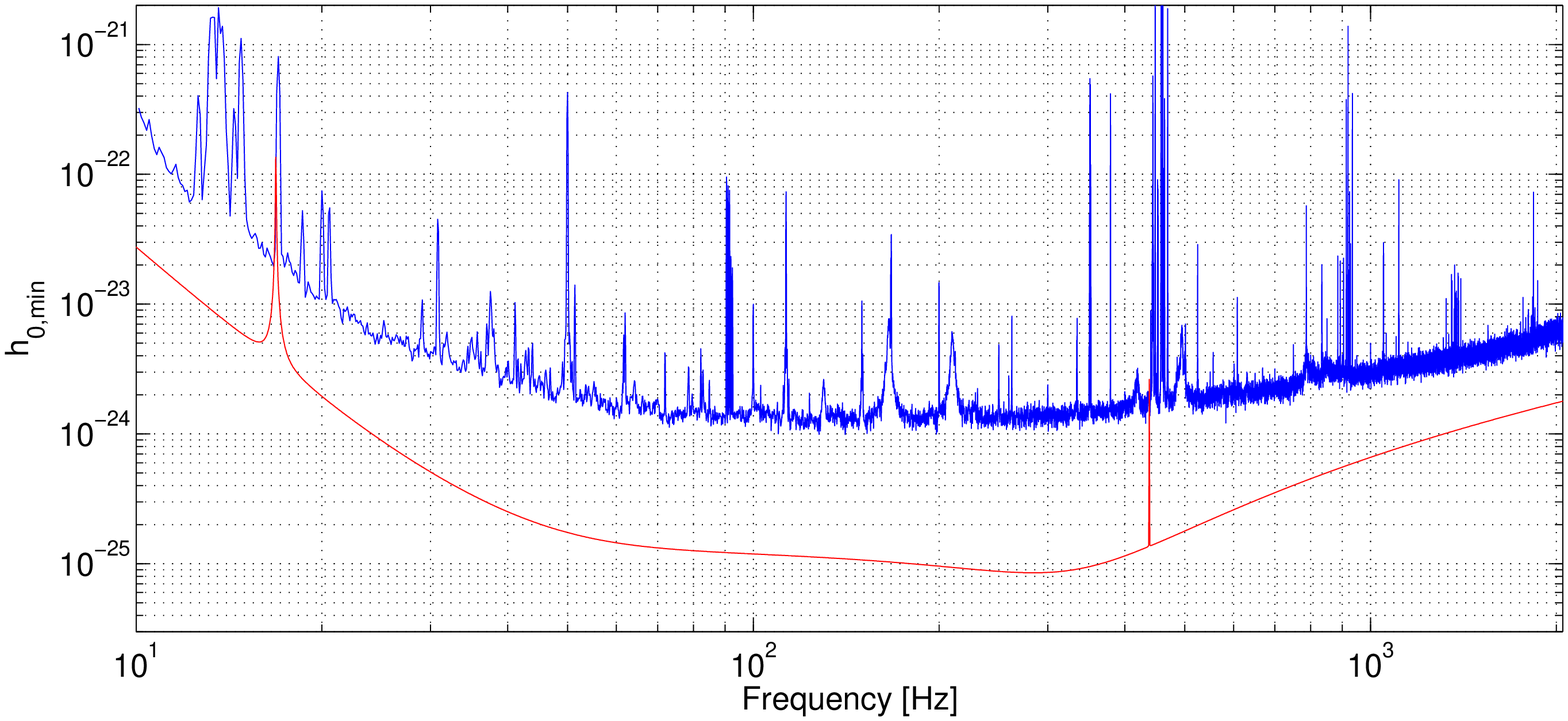} 
\includegraphics[width=10cm]{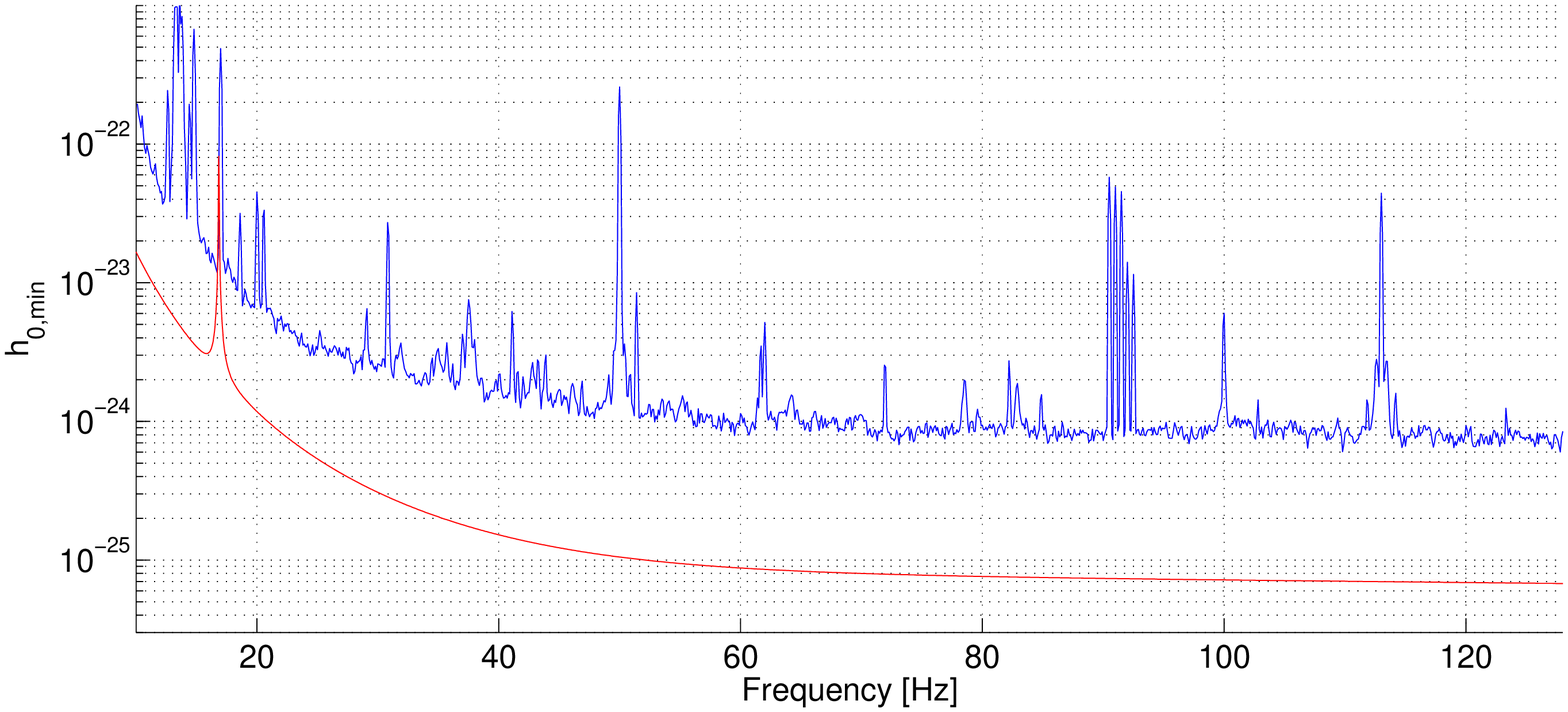} 
\caption{95$\%$ confidence level sensitivity curves computed using Eq.(\ref{h0min}) and taking a typical Virgo VSR4 sensitivity curve (upper blue curve) and the planned Advanced Virgo sensitivity curve (lower red curve) with $T_{obs}=1$ year. Upper plot has been obtained using  $T_{FFT}=1024$ seconds and assuming to select $10^9$ candidates, while the bottom plot, which refers to the frequency range 10-128 Hz, has been obtained taking $T_{FFT}=8192$ seconds and assuming to select $10^7$ candidates. In both cases $\tau_{min}=1000$ years is used.}\label{fig:sens_1year}
\end{figure*}

The sensitivity, at a given confidence level $\Gamma$ (e.g. $95\%$), is defined as
the minimum signal amplitude which would produce a candidate in a fraction $\ge
\Gamma$ of a large number of repeated experiments. It does not depend on the actual
result of the analysis. To compute it we start from the expression for the
probability of selecting a candidate as a function of the signal amplitude and impose
that it is equal to $\Gamma$:  
\be
P_{n>n_{thr}}(\lambda)=\int_{n_{thr}}^{\infty} P_n(\theta_{thr};\lambda)dn=\Gamma
\ee
Using the Gaussian approximation we have
\be
\erfc{\left(\frac{n_{thr}-Np_{\lambda}}{\sqrt{2Np_{\lambda}(1-p_{\lambda})}}\right)}=2\Gamma
\label{pdet}
\ee
Note that this equation can be obtained from Eq.(\ref{pfd2}) by putting
$\Gamma=1-P_{fd}$ and using the identity $\erfc{(x)}=2-\erfc{(-x)}$.
From the previous equation, using Eq.(\ref{nthr_gauss}), together with
Eq.(\ref{pfa_cand}), we can write
\begin{widetext}
\be
Np_0+\sqrt{2Np_0(1-p_0)}{\erfc}^{-1}(2\frac{N_{cand}}{N_{tot}})-Np_{\lambda}-\sqrt{2Np_{\lambda}(1-p_{\lambda})}\erfc^{-1}(2\Gamma)=0
\label{pdet2}
\ee
\end{widetext}
Solving this equation using the small signal approximation of Eq.(\ref{eq:plambda_approx}), as discussed in Appendix \ref{senseval}, we find the minimum detectable spectral amplitude
\begin{equation}
\lambda_{min}\approx
\frac{2}{\theta_{thr}}\sqrt{\frac{p_0(1-p_0)}{Np^2_1}}\left(CR_{thr}-\sqrt{2}\erfc^{-1}(2\Gamma)\right)=\frac{\Lambda_1}{\sqrt{N}}
\label{lambdamin}
\end{equation}
where $p_1=e^{-\theta_{thr}}-2e^{-2\theta_{thr}}+e^{-3\theta_{thr}}$.
The coefficient $\Lambda_1$ is given in Tab.(\ref{thres}) for various choices of
the search parameters.
As shown in Appendix \ref{senseval}, the minimum detectable spectral amplitude of Eq.(\ref{lambdamin}) corresponds to a
minimum detectable strain amplitude $h_{0,min}$ given by
\begin{widetext}
\be
h_{0,min}\approx
\frac{4.02}{N^{1/4}\theta^{1/2}_{thr}}\sqrt{\frac{S_n(f)}{T_{FFT}}}\left(\frac{p_0(1-p_0)}{p^2_1}\right)^{1/4}\sqrt{CR_{thr}-\sqrt{2}\erfc^{-1}(2\Gamma)}=\frac{\Lambda}{N^{1/4}}
\sqrt{\frac{S_n(f)}{T_{FFT}}}
\label{h0min}
\ee 
\end{widetext}
The coefficient $\Lambda$ is given in Tab.(\ref{thres}) for different values of the
search parameters.
By inverting Eq.(\ref{h0min}), at fixed false alarm - i.e. fixed $CR_{thr}$ - we can
express the detection probability $\Gamma$ as a function of the sensitivity:
\be
\Gamma=\frac{1}{2}\erfc{\left[\frac{1}{\sqrt{2}}\left(CR_{thr}-\frac{h^2_{0,min}}{\beta^2}\right)\right]}
\ee
where
\be
\beta=\frac{4.02}{N^{1/4}\theta^{1/2}_{thr}}\sqrt{\frac{S_n(f)}{T_{FFT}}}\left(\frac{1-p_0}{p_0}\right)^{1/4}
\ee
The 95$\%$ confidence level sensitivities are plotted in Fig.(\ref{fig:sens_1year}) using a typical Virgo VSR4 run sensitivity curve ($T_{obs}\simeq 90$ days) and the planned Advanced Virgo sensitivity curve (assuming $T_{obs}=1$ year). Upper plot covers the range 10-2048 Hz and has been obtained taking $T_{FFT}=1024$ s and assuming to select $10^9$ candidates, while the bottom plot, which refers to the frequency range 10-128 Hz, has been obtained using $T_{FFT}=8192$ seconds and assuming to select $10^7$ candidates. In both cases $\tau_{min}=1000$ years has been taken.

\subsection{Sensitivity loss with respect to the optimal method}
\label{comparison}
The optimal method to search for a monochromatic signal of unknown frequency
consists in computing an estimation of the power spectrum, e.g. by means of the
periodogram, and searching for statistically significant peaks. By making a
periodogram of length $T_{obs}$ we have an amplitude SNR 
\be
SNR_{opt}=\frac{h_0}{2}\sqrt{\frac{T_{obs}}{S_n(f)}}
\label{snr_opt}
\ee  
The corresponding {\it nominal} sensitivity (i.e. corresponding to SNR=1) is
\be
\bar{h}_{0}=2\sqrt{\frac{S_n(f)}{T_{obs}}}
\label{h0nom}
\ee
The most basic incoherent combination of the data consists in dividing
the observation period in $M$ pieces and compute the spectrum for each and then sum.
In this case we have, see e.g. \cite{astold}
\be
SNR_{M}=\frac{h_0}{2}\sqrt{\frac{T_{obs}}{S_n(f)\sqrt{M}}}=\frac{SNR_{opt}}{M^{1/4}}
\label{snr_n}
\ee
The corresponding {\it nominal} sensitivity is
\be
\bar{h}_{0,M}=\bar{h}_{0}M^{1/4}
\label{h0n}
\ee
For instance, for $T_{obs}=1~yr$ and $T_{FFT}=1024~s$ we have $M=61594$ (assuming the FFT are
interlaced by half). This implies a nominal sensitivity loss of the incoherent combination with respect to the optimal analysis of $(61594)^{1/4}=15.7$. 

In practice, when a wide-parameter search is done even in the case of an optimal analysis we need anyway to set a threshold
to select a reasonable number of candidates. As the spectral power is distributed
exponentially, the probability of having in a given frequency bin a power $S$ larger
than a threshold $S_{thr}$ is 
\be
P(S>S_{thr})=e^{-S_{thr}}
\ee
If we impose that the number of candidates above $S_{thr}$ is $N_{cand}$, then we
have $N_{tot}\cdot e^{-S_{thr}}=N_{cand}$ and
\be
S_{thr}=-\log\left(\frac{N_{cand}}{N_{tot}}\right)
\ee
where $N_{tot}$ is the total number of points in the source parameter space. For
instance, taking $\delta t=2.44\cdot 10^{-4} ~s$, $T_{obs}=1~yr$ and
$\tau_{min}=10^3~yr$ we find
$j_{max}=3$ and $N_{tot}\simeq 8.20\cdot 10^{40}$ and the threshold we
should choose is
\be
S_{thr}=-\log\left(\frac{10^9}{8.2\cdot 10^{40}}\right) =73.5
\ee

The spectrum distribution in presence of a signal of amplitude $\lambda$ is a non-central $\chi ^2$ with
two degrees of freedom, see Eq.(\ref{noncenchi2}). The probability of having a value of the spectrum, in a given
frequency bin, larger than a threshold $S_{thr}$ is then
\be
P(S>S_{thr};\lambda)=\int_{S_{thr}}^{\infty}
e^{-S-\frac{\lambda}{2}}I_0\left(\sqrt{2S\lambda}\right)
\ee
This is the detection probability. We can compute the sensitivity by determining
that value of signal amplitude, $\lambda_{min}$, such that the detection probability
is, e.g., $\Gamma=0.95$. This can be done numerically. For instance for
$\Gamma=0.95$ we find $\lambda_{min}=188.4$. 
In order to compare this optimal sensitivity to the Hough transform sensitivity it
is more convenient to work with $h_0$ instead of $\lambda$. In terms  of averaged
$h_0$ the optimal sensitivity, assuming to select $10^9$ candidates, can be written as 
\be
h_{0,opt}\simeq 39\sqrt{\frac{S_n}{T_{obs}}}
\ee
while the Hough sensitivity is given by Eq.(\ref{h0min}). The ratio of the latter to
the former is, for $\Gamma=0.95$
\be
R\simeq\frac{\Lambda}{39N_{FFT}^{1/4}}\sqrt{\frac{T_{obs}}{T_{FFT}}}\simeq \frac{\Lambda}{46.4}\left(\frac{T_{obs}}{T_{FFT}}\right)^{1/4}
\ee
Taking, e.g., $T_{obs}=1~yr$, $T_{FFT}=1024~s$ and $\tau=1,000~yr$ the ratio is
$R\sim 3.7$, while it is about 2.3 for $T_{FFT}=8192$ seconds. Then, even if the nominal sensitivity loss can be large, the actual loss, by taking into account the need to select a given number of candidates, is much smaller. This is due to the different probability distribution of the quantities over which candidates are selected, power spectrum for the optimal analysis, critical ratio (or number count) for the Hough transform.

\section{\label{pmclean} Removal of time and frequency domain disturbances}
The presence of time and frequency domain disturbances in detector data affects
the search and, if they are not properly removed, reduces the search sensitivity or even blinds the search
at given times and/or in given frequency bands. The effect in the analysis 
varies, depending on their nature and on their amplitude.  
It is therefore very important to apply procedures to safely 
remove them or reduce their effect, without 
contaminating a possible CW signal.
The disturbances can be catalogued as ``time domain glitches'', which enhance
the noise level of the detector in a wide frequency band, 
``spectral lines of constant frequency'', 
sometimes of known origin, like calibration lines or lines whose
origin has been discovered by studying the behaviour of the detector and the surrounding environment, and ``spectral wandering lines'', where the frequency of the disturbance 
moves in time, which are typically of unknown origin and might 
be present only for a few days or even hours. Moreover, simulated signals from spinning neutron stars are injected in the detector for testing purposes (hardware injections). To do real analyses, however, these signals have to be removed, 
as they are clear ``artifacts'' and in some cases so huge that the discovery
of real GW signals around the frequency of the injection could be impossible.

Different kind of disturbances are identified and removed by using different techniques, 
which are applied in different steps of the analysis, as described below.

\subsection{Removal of time domain glitches}
Time domain glitches are identified and removed during the construction of
the SFDB. This kind of disturbances shows up randomly and enhances the noise level in a wide frequency band.
The size of the affected frequency band depends on the structure of 
the glitch.
The procedure we apply has been described in \cite{Cleaning} and is only summarized here. 
We identify big glitches by the application of a high-pass bilateral 
filter to the data. The filter is bilateral as the high-passed data have to be
in phase with the original data. The cutoff of the high-pass filter depends
of the maximum frequency of the FFTs we are constructing 
(e.g. it can be 100 Hz for the 1024 s FFTs, whose maximum frequency is 2048 Hz) 
We then {\it subtract} these glitches from the 
original time series, with the advantage of not reducing the observation time,
an important requirement for CW searches, and of not substituting 
the data with zeroes, which would cause an evident loss of any information present in them.
It is not possible to quantify the overall effect of this cleaning in a general way, as 
the actual improvement depends on the characteristics of the detector and of the 
specific dataset considered. But we consider this procedure important in any case: depending on
the situation the final effect will be more or less relevant but, as a basic principle,
it is important to remove these artifacts by maintaining the information in the data and avoiding 
to reduce the observing time, which affects the final sensitivity of the search.
References \cite{Cleaning} and \cite{Cleaningpaola} describe two opposite situations: a very big improvement in sensitivity in one case, and a nearly null improvement in the other case.

\subsection{Removal of spectral wandering lines}
\begin{figure*} [hbtp]  
\includegraphics[width=10cm]{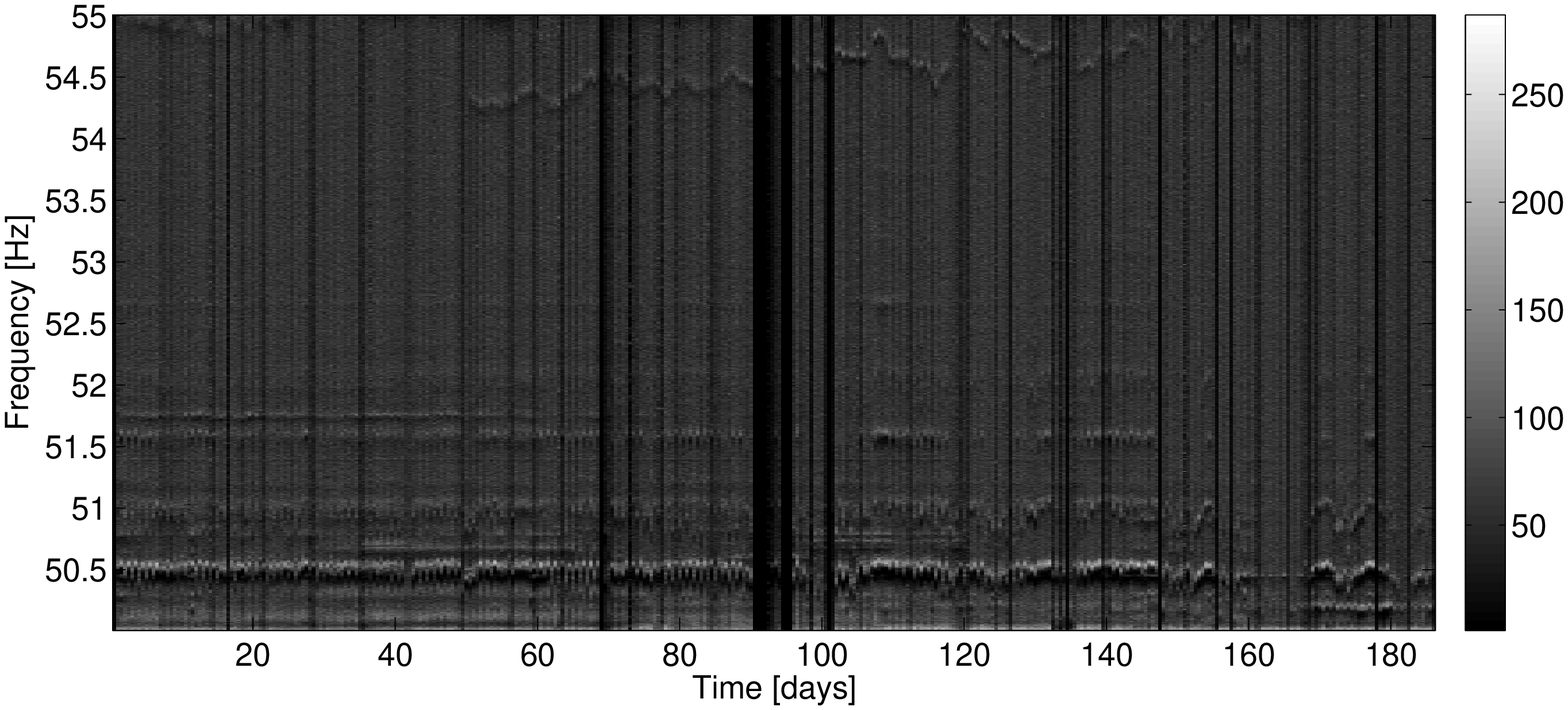}
\includegraphics[width=10cm]{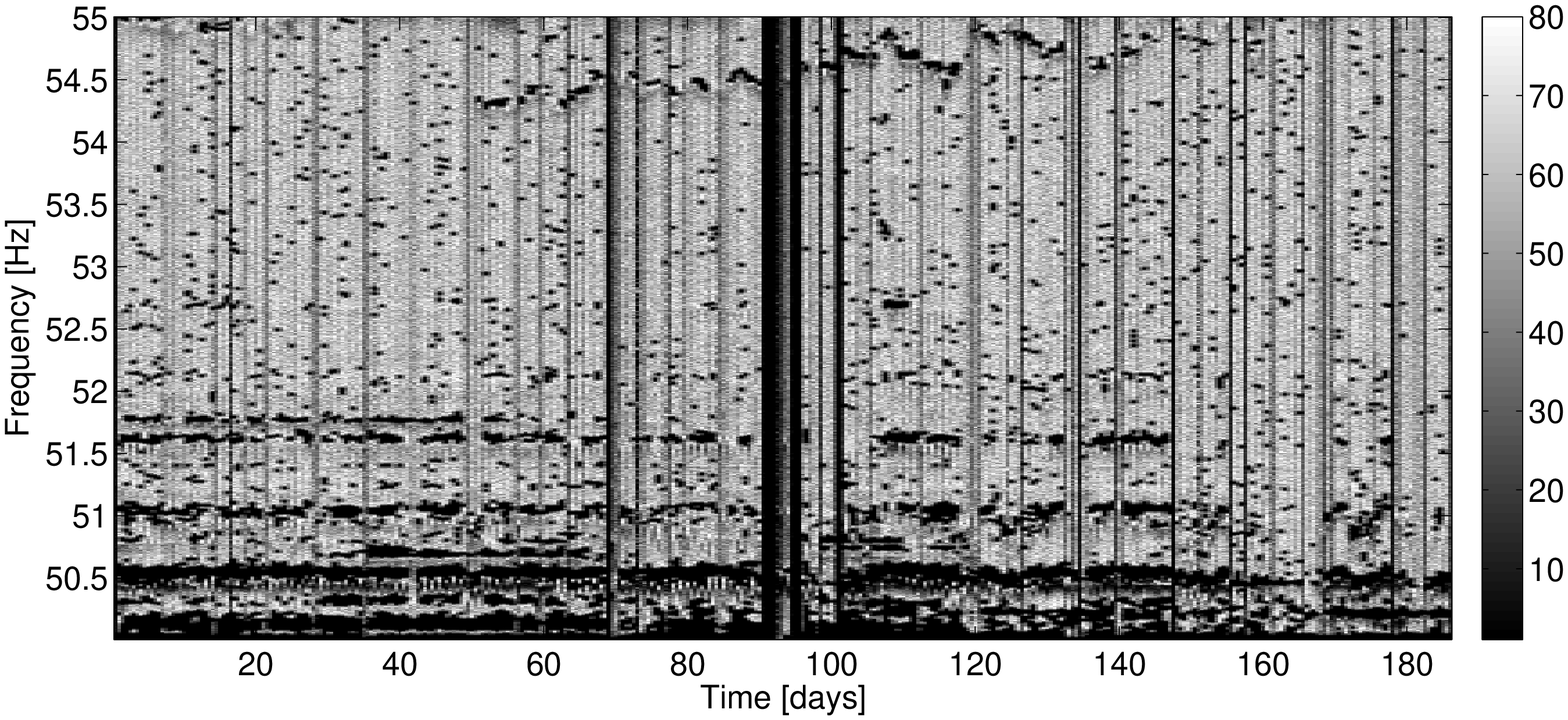}
\caption{An example of the removal of spectral wandering lines on VSR2 
data, in the frequency range [50-55] Hz.
The upper plot is the gross histogram of the peakmap while the bottom plot is the same after the
cleaning procedure, having put the threshold for the veto to 80. Here the darkest spots correspond
to the removed bins, and in fact the removal of the wandering line around 54.2-54.5 Hz 
is well evident being tracked by a dark path.
The two histograms have different scales on the z-axis, as the maximum value after the cleaning is 80.  
}  
\label{fig:CleanPeak}
\end{figure*}

Noise spectral lines present in the FFTs, if strong enough to be local maxima
of the equalized spectrum, are selected by the procedure which 
constructs the peakmaps and if persistent enough 
their final effect in the Hough analysis and in the extraction of the candidates can be dramatic. 
If the frequency of these disturbances changes with time, 
either randomly or according to some rule, it is clearly not optimal to veto the whole band affected by the line,
as this would imply the removal of too many data. It is therefore important to develop a 
method which is able to remove only the time and frequency bins of the peakmap really touched
by the noise line.
The idea is to construct an histogram of a low resolution (both in time and in frequency) peakmap, which we call ``gross histogram''.
The choice for the gross resolution is mainly made by considering the possible presence of a CW signal, which must have a completely negligible effect after 
the integration (as we do not want to remove it) and 
a reasonable time extent for detector non-stationarities.
Over a time scale of the order of one day or less the Doppler effect which matters is only that due to
the Earth rotation, which size is 
$f_0  \frac{\Omega_{rot} R_{\oplus}}{c}$ Hz, being $\Omega_{rot}$ the Earth rotation angular frequency and $R_{\oplus}$ the Earth radius, 
which gives, for example, $\sim 1.9\times 10^{-4}$ Hz at a frequency of 128 Hz. A possible reasonable choice could be 
$\Delta t_H$ = 12 hours for time resolution and $\Delta f_H$ =0.01 Hz for frequency resolution. In this way any real CW signal would be completely confined within one bin and would not significantly contribute to the histogram.
The choice of the threshold to veto the (gross) bins containing artifacts can be done with
the following reasoning.
The distribution of the average noise in each peakmap is binomial with parameter 
$p_0 \simeq 0.0755$, as shown in Sec.\ref{sec:peakmap}, and the expected value 
in a single gross bin is $E_H=N \times p_0$, 
where $N=\frac{\Delta t_H}{T_{FFT}}\cdot \frac{\Delta f_H}{\delta f}$ is the number of ``points'' in it. For FFTs interlaced by the half it can be also written as $2\Delta t_H\cdot \Delta f_H$.
Thus the expected value and the standard deviation 
do not depend on the frequency band considered, at least ideally, and the threshold to
veto artifacts can be fixed  on the basis of the value of $N$.
As an example, for (overlapping by the half) FFTs of duration 8192 s and considering $\Delta t_H=12$ hours, 
$\Delta f_H=0.01$ Hz we have  
$N\simeq 864$. The expected value of the distribution is thus $E_H=65$
and the standard deviation $\sigma_H=\sqrt{N\,p_0\,(1-p_0)}$ = 7.8.
A reasonable choice for the threshold can be given by 2-3 standard deviations from the expected value, which in the example given corresponds to the range [80-88]. The gross bins with amplitude above the threshold are removed from the peakmap before applying the FH transform.
Figure \ref{fig:CleanPeak} shows an example of the cleaning effect on VSR2 
data in the frequency range [50-55] Hz.
The left plot is the gross histogram of the peakmap and the right plot is the same after the
cleaning procedure, having put the threshold for the veto to 80. 
The presence of a wandering line, which roughly moves from 52.2 to 52.5 Hz is evident in 
the upper plot and the fact it has been removed by the cleaning is visible in the bottom plot by
following the dark track. 
\subsection{Removal of spectral lines of constant frequency}
The presence of spectral lines of constant frequency would also affect the Hough analysis and
the candidates selection, but in this case the removal of the disturbed bins in the peakmap
is much simpler than in the previous case, as here the affected frequency band is very small.
Different similar procedures can be used for this. 

One possibility is to use the list of known lines and remove all the frequency bins in the list. 
This is very simple and fast, but usually the frequency resolution used for detector 
characterization is worse compared to the one of typical CW analyses. 
As a consequence by using this method the frequency bins removed will be more than what really
needed. 

Another possibility is to run a ``persistency'' analysis on the peakmaps, by histogramming
the frequency bins and deciding a proper threshold on the basis of some statistical properties of
the histograms. We can use the average and  standard deviation of the number count to decide the
threshold or we can use a better ``robust'' statistic, as the one described in Appendix \ref{ROBUST}, which is based on the median rather than the mean, much less affected by tails in the distribution.
All the frequency bins exceeding the chosen threshold are 
then removed from the analysis. The advantage here is that this 
is also quite simple and the frequency resolution is the same used 
in the analysis.

Still another option is based on the Hough procedure that can be used to identify these 
disturbances, by computing it on the peakmaps without any Doppler 
correction and looking for spin-down values 
around zero, e.g. $\dot{f}=0 \pm \delta \dot{f}$, 
to admit some variation in the frequency of the artifact.
This has the advantage of identifying disturbances which would show up 
in the real analysis, reducing the amount of manipulation on the data. 

To correctly use the last two procedures the comparison of all the removed lines with
the list of known disturbances produced by detector characterization studies
(e.g. done in Virgo using the NoEMi tool \cite{noemi}) is mandatory.  
In fact true CW signals with negligible 
Doppler modulation, as those near the ecliptic poles, 
might be seen as disturbances here. 
Thus, all the lines removed by the procedure and not in the 
known lines list will then be studied to understand if they were due to the instrument
and, if not, analyzed, by removing the veto.

\subsection{Removal of hardware injections}
The removal of the hardware injections possibly present in the band to be analyzed,
is another very important step of the procedure.
Given the fact that the parameters of these signals are known, their removal from the peakmap is pretty simple
and precise: we can remove the exact bins where the injections were done, by using the known frequency, spin-down and by evaluating the Doppler effect.
In Figure \ref{fig:peakHI3} we have already shown the peakmap of VSR2 Virgo data
around one hardware injection. The procedure is 
designed to remove the signal bins, which are the dark blue dots. The fact that in this example the
signal is so huge that the track is visible with naked eyes does not 
mean anything for the removal procedure, which is based only on the known 
information about the injected parameters and not on the signal amplitude.\\

\section{\label{conc}Conclusions}
In this paper we have described a new hierarchical analysis method for the all-sky search of continuous gravitational wave signals. Particular attention has been put in properly taking into account issues related to the use of real data and the computational aspects. Several novelties with respect to other similar methods are discussed. The core of the pipeline is the frequency-Hough transform, a particularly efficient implementation of the Hough transform, which is used as incoherent step of the analysis. Both a coarse and a refined grid in the parameter space are used to select the candidates. The coarse grid is heavily over-resolved in frequency without increasing the computational load of the analysis, thanks to the features of the FH. This allows to significantly reduce the sensitivity loss associated to digitization. Coarse candidates are selected in a way to minimize the blinding effect of disturbances present in the data. Once coarse candidates have been chosen, a refined analysis using over-resolution also in sky position and spin-down is performed only around them, still using the FH. This allows to reduce the uncertainty in candidate parameters, which is crucial for the coincidence step. In fact, in order to reduce the false alarm probability given the sets of candidates found in the analysis of two (or more) datasets belonging to different runs of the same detector or to different detectors, coincidences are done and only surviving candidates are furtherly processed. Having a more accurate determination of candidate parameters implies a smaller number of surviving candidates after coincidences. Coincidences are preceeded by a clusterization step in which nearby candidates are grouped together. Coincident candidates are subject to a verification step with the aim of discarding them or significantly improve the detection efficiency. Finally on remaining candidates a follow-up with longer coherence time is applied, which allows to increase the signal-to-noise ratio of detected signals and to better estimate their paramters. 
Moreover, several data cleaning procedures are applied in order to remove noise disturbances and then improve the search sensitivity.
First, the removal of short duration time domain glitches is done before constructing the SFDB. Then, three further cleaning steps are applied at the level of the peakmaps. Wandering spectral lines are carefully removed by using a low-resolution histogram of the peakmap. Three alternative methods to identify and cancel spectral lines of constant frequency are presented. Hardware injected signals are also removed bin by bin in the peakmap. 

In the immediate we plan to apply this method to the analysis of Virgo VSR2 and VSR4 data. In particular, the low frequency sensitivity of these data is significantly better than that of LIGO data over which wide-parameter searches of CW have been concentrated so far. More in the future we will use it to analyze data from advanced Virgo and LIGO detectors, which will start their science runs in 2015-2016.

\begin{acknowledgments}
We want to thank LSC-Virgo Continuous Waves group for the useful discussions and the anonymous referees for constructive comments that allowed us to improve the paper.
\end{acknowledgments}

\appendix

\section{Theory of coincidences}
\label{sec:theory_coin}

In this section we derive the number of expected coincidences among two sets of candidates, in the hypothesis of Gaussian noise.
In any realistic case each candidate must be associated with a {\it coincidence window}, that is a small volume of the parameter space around the candidate parameters. Two candidates belonging to two different sets are coincident when they coincidence windows overlap. 

Let us then assume to have two sets of candidates, belonging to two different detectors or two different runs of the same detector, with $N_1$ and $N_2$ elements respectively. Each candidate is completely defined by the values of $M$ parameters. in the all-sky search described in this paper we have $M=4$ (position, frequency, first order spin-down). Let us indicate with $m_{i;j}$ the number of values the $j$th parameter can assume for candidates of the $i$th set, i.e. the number of cell (e.g. the number of frequency bins or the number of different spin-down values). Let us assume that, with respect to a given parameter, the candidates are distributed uniformly. Then, the probability of having $k$ candidates, among those of the $i$th set, in a given cell of the $j$th parameter is given by a Poisson distribution:
\begin{equation}
P(k;\mu_{i;j})=\frac{\mu^k_{i;j}}{k!}e^{-\mu_{i;j}}
\label{eq:pois}
\end{equation}
where 
\begin{equation}
\mu_{i;j}=\frac{N_i}{m_{i;j}}
\end{equation}
The probability of having 0 candidates in a given cell is 
\begin{equation}
P_0=P(0;\mu_{i;j})=e^{-\mu_{i;j}}
\end{equation}
Hence, the probability of having {\it at least} $k$ consecutive empty cells is 
\begin{equation}
P_{0,k}=P^k_0=e^{-k\mu_{i;j}}
\end{equation}  
Let us now consider a coincidence window (symmetric with respect to the central value) $w_{i;j}=2n_{i;j}+1$, expressed as a number of cells and introduce the total coincidence window
$w_j=2(n_{1;j}+n_{2;j})+1$. Given a candidate of the first set, if it does not coincide, with respect to the parameter $j$th, with a candidate of the second set within the window $w_j$ this means that between the two nearest candidates of the second set there must be at least $w_j$ empty cells and this has probability 
\begin{equation}
P^{w_j}_0=e^{-w_j\cdot \mu_{2;j}}
\end{equation}
assuming the grid step is the same.  Hence, the probability that a candidate in the first set is coincident with at least one candidate of the second set is
\begin{equation}
P_{1\rightarrow 2}=1- e^{-w_j\cdot \mu_{2;j}}
\end{equation}
Then, the expect number of coincidence of the candidate of the first set with at least one candidate of the second set is 
\begin{equation}
N_{coinc,1}=N_1\cdot P_{1\rightarrow 2}
\end{equation}
Similarly, the expected number of candidates of the second set in coincidence with at least one candidate of the first set is
\begin{equation}
N_{coinc,2}=N_2\cdot P_{2\rightarrow 1}
\end{equation}

These relations can be easily generalized to the case in which coincidences are done among more than one parameter. In the general case the total number of cells in the parameter space for candidates of the $i$th set is 
\begin{equation}
m_i=\prod_{j=1}^{M}m_{i;j}
\end{equation}
The expected number of coincidences is given by 
\begin{equation}
N_{coinc,1}=N_1\left(1-e^{-N_2\prod_{j=1}^{M}\frac{w_j}{m_{1;j}}}\right)
\end{equation}
 \begin{equation}
N_{coinc,2}=N_2\left(1-e^{-N_1\prod_{j=1}^{M}\frac{w_j}{m_{2;j}}}\right)
\end{equation}
If $\mu_{i;j}\ll 1$, which will be well satisfied in general, we can use the approximation $e^{-\alpha x}\approx 1-\alpha x$ to obtain
\begin{equation}
N_{coinc,1}\approx N_1\cdot N_2\prod_{j=1}^{M}\frac{w_j}{m_{1;j}}
\end{equation}
\begin{equation}
N_{coinc,2}\approx N_1\cdot N_2\prod_{j=1}^{M}\frac{w_j}{m_{2;j}}
\end{equation}
If the grid in the parameter space is the same for both candidate sets, i.e. $m_1=m_2=m$, then we have 
\begin{equation}
N_{coinc}=N_{coinc,1}=N_{coinc,2}\approx N_1\cdot N_2\prod_{j=1}^{M}\frac{w_j}{m_{j}}
\label{eq:ncoin_approx}
\end{equation}
In the ideal case in which no window is used, i.e. $w_j=1$ for all $j$s, we would have simply
\begin{equation}
N_{coinc}\approx \frac{N_1\cdot N_2}{m}
\end{equation}
where $m=\prod_{j=1}^{M}m_j$ is the total number of cells in the parameter space. 
\section{Sensitivity evalutation}
\label{senseval}
Using the small signal approximation, Eq.(\ref{eq:plambda_approx}), and neglecting terms
of order $o(\lambda^2)$ and higher Eq.(\ref{pdet2}) can be written as
\begin{widetext}
\be
\sqrt{2Np_0(1-p_0)}{\erfc}^{-1}(2\frac{N_{cand}}{N_{tot}})-Np_1\theta_{thr}\frac{\lambda}{2}-\sqrt{2N\left[p_0(1-p_0)+p_1\theta_{thr}\frac{\lambda}{2}(1-2p_0)\right]}\erfc^{-1}(2\Gamma)=0
\label{pdet3}
\ee
\end{widetext}
Equation \ref{pdet3} can be put in the form
\be
(A+B\lambda)^2=C+D\lambda
\label{pdet4}
\ee
where
\begin{align}
A &= \sqrt{2Np_0(1-p_0)}{\erfc}^{-1}(2\frac{N_{cand}}{N_{tot}}) \notag \\
B &= -Np_1\frac{\theta_{thr}}{2} \notag \\
C &= 2Np_0(1-p_0)\left(\erfc^{-1}(2\Gamma)\right)^2 \notag \\
D &= 2Np_1(1-2p_0)\frac{\theta_{thr}}{2}\left(\erfc^{-1}(2\Gamma)\right)^2
\end{align}
By writing Eq.(\ref{pdet4}) as
\be
B^2\lambda^2+(2AB-D)\lambda+A^2-C=0
\ee
we can write the solution as
\be
\lambda=\frac{-2AB+D\pm \sqrt{\Delta}}{2B^2}
\ee
where
\be
\Delta=D^2-4ABD+4B^2C
\ee
For any reasonable value of $N$, of $N_{cand}$ and $N_{tot}$, and given the values
of $p_0$ and $\theta_{thr}$ it comes out that $D\ll AB$, then the solution can be
written as
\be
\lambda \approx -\frac{A}{B}\pm \sqrt{\frac{-AD+BC}{B^3}}
\label{pdet5}
\ee
We take as physical solution that having the minus sign, because the minimum
detectable amplitude must increase as $\Gamma$ becomes larger than $0.5$
($\erfc^{-1}(2\Gamma) =0$ for $\Gamma=0.5$ and becomes negative for $\Gamma>0.5$):
\begin{widetext}
\begin{equation}
\lambda_{min}\approx
\frac{\sqrt{2Np_0(1-p_0)}{\erfc}^{-1}(2\frac{N_{cand}}{N_{tot}})}{Np_1\frac{\theta_{thr}}{2}}-
2\sqrt{2}\frac{\erfc^{-1}{(2\Gamma)}}{Np_1\theta_{thr}}\sqrt{\sqrt{2Np_0(1-p_0)}(1-2p_0)\erfc^{-1}(2\frac{N_{cand}}{N_{tot}})+Np_0(1-p_0)}
\label{pdet6}
\end{equation}
\end{widetext}
As for typical values of $N$ we have $\sqrt{2Np_0(1-p_0)}(1-2p_0)\erfc^{-1}(2\frac{N_{cand}}{N_{tot}})\ll 2Np_0(1-p_0)$, then 
\begin{widetext}
\begin{equation}
\lambda_{min}\approx
\frac{2\sqrt{2}}{\theta_{thr}}\sqrt{\frac{p_0(1-p_0)}{Np^2_1}}\left(\erfc^{-1}(2\frac{N_{cand}}{N_{tot}})-\erfc^{-1}(2\Gamma)\right)=
\frac{2}{\theta_{thr}}\sqrt{\frac{p_0(1-p_0)}{Np^2_1}}\left(CR_{thr}-\sqrt{2}\erfc^{-1}(2\Gamma)\right)
\label{lambdamin2}
\end{equation}
\end{widetext}

Now we
want to express the sensitivity in terms of the minimum detectable strain amplitude, $h_{0,min}$. We follow here the discussion in \cite{hough1}. 
The GW signal can be written as:
\begin{align}
h(t) &= F_+h_{0+}\cos{(\phi(t))}+F_{\times}h_{0\times}\sin{(\phi(t))} \notag \\
h_{0+} &= h_0\frac{1+\cos^2 \iota}{2} \notag \\
h_{0\times} &= h_0\cos \iota
\label{hoft}
\end{align}
In order to arrive to an expression for the minimum detectable strain amplitude,
$h_{0,min}$ we explicitly compute the signal Fourier transform that appears in
Eq.(\ref{lambdadef}) and then average over the various parameters. The Fourier
transforms of the sine and cosine with frequency $f_0$ are
\begin{align}
Y_1(f) &= \frac{\left(\delta(f-f_0)+\delta(f+f_0)\right)}{2} \notag \\
Y_2(f) &= -j\frac{\left(\delta(f-f_0)-\delta(f+f_0)\right)}{2} 
\end{align}
In a time $T_{FFT}$ the signal frequency does not shift by more than half a
frequency bin, by construction, and can be considered roughly constant. The Fourier
transform of a finite length signal is the convolution of the Fourier transform of
the signal with the Fourier transform of a rectangular window of length $T_{FFT}$:
\be
Z(f)=\int_{-\infty}^{+\infty} Y(f')\cdot \frac{\sin{\pi(f-f')T_{FFT}}}{\pi(f-f')}df'
\ee
In the case of our sinusoidal signals, by taking as the signal frequency the
frequency $f_k$ at the bin center and considering only positive frequencies we have
\begin{align}
Z_1(f) &= \frac{h_0}{2}\frac{\sin{\pi(f-f_k)T_{FFT}}}{\pi(f-f_k)} \notag \\
Z_2(f) &= -j\frac{h_0}{2}\frac{\sin{\pi(f-f_k)T_{FFT}}}{\pi(f-f_k)} 
\end{align}
So the Fourier transform of the finite length GW signal given by Eq.(\ref{hoft}) is
\be
\tilde{h}(f)\approx T_{FFT}\frac{\left(F_+A_+-jF_{\times}A_{\times}\right)}{2}
\frac{\sin{\pi(f-f_k)T_{FFT}}}{\pi(f-f_k)T_{FFT}} 
\label{hoftfour}
\ee
The square modulus is:
\be
|\tilde{h}(f)|^2\approx
T^2_{FFT}\frac{\left(F_+A_++F_{\times}A_{\times}\right)^2}{4}\left(\frac{\sin{\pi(f-f_k)T_{FFT}}}{\pi(f-f_k)T_{FFT}}\right)^2
\label{htilde2}
\ee
We now take the average of all varying quantities. The two beam pattern functions
$F_+,~F_{\times}$ depend on the source position and wave polarization angle. It easy
to verify that
\begin{align}
<F^2_+>_{\alpha,\delta,\psi}=<F^2_{\times}>_{\alpha,\delta,\psi}=\frac{1}{5} \notag \\
<F_+\cdot F_{\times}>_{\alpha,\delta,\psi}=0
\end{align}
The two amplitudes $A_+,~A_{\times}$ depend on the angle $\iota$ between the star
rotation axis and the line of sight and 
\be
<A^2_++A^2_{\times}>_{\cos{\iota}}=\frac{1}{2}\int_{-1}^1
\left[\left(\frac{1+\cos^2{\iota}}{2}\right)^2+\cos^2{\iota}\right]d\cos{\iota}=\frac{4}{5}
\ee
We average the frequency dependent part of Eq.(\ref{htilde2}) over $f$ in the range
$[f_k-\frac{\delta f}{2},f_k+\frac{\delta f}{2}]$. By changing variable,
$x=\pi(f-f_k)T_{FFT}$, we have
\be
\frac{1}{\delta f}\int_{f_k-\frac{\delta f}{2}}^{f_k+\frac{\delta f}{2}}
\left(\frac{\sin{(\pi(f-f_k)T_{FFT})}}{\pi(f-f_k)T_{FFT}}\right)^2df =
\frac{1}{\pi}\int_{-\pi/2}^{\pi/2}\frac{\sin^2{x}}{x^2}dx=\frac{2.4308}{\pi}
\ee
In terms of the signal spectral amplitude we can then write:
\be
<\lambda>_{\alpha,\delta,\psi,f}\approx\frac{4h^2_0}{S_n(f)}\frac{2.4308}{25\pi}T_{FFT}
\label{lambda_ave}
\ee
We now equate Eq.(\ref{lambda_ave}) to Eq.(\ref{lambdamin2}) finding the minimum
signal amplitude $h_{0,min}$ that would produce a candidate in the Hough map:
\begin{widetext}
\be
h_{0,min}\approx
\frac{4.02}{N^{1/4}\theta^{1/2}_{thr}}\sqrt{\frac{S_n(f)}{T_{FFT}}}\left(\frac{p_0(1-p_0)}{p^2_1}\right)^{1/4}\sqrt{CR_{thr}-\sqrt{2}\erfc^{-1}(2\Gamma)}
\sqrt{\frac{S_n(f)}{T_{FFT}}}
\label{h0min_app}
\ee 
\end{widetext}

\section{Construction of the coarse grid in the sky}
\label{A2}
We give here some detail on the practical construction of the coarse grid in the sky, described in Sec.\ref{grid}.
We start from a set of 3 points at different ecliptic latitude, the poles and the equator:
$$ \beta_1= \pi/2, \beta_N=-\pi/2, \beta_{\frac{N+1}{2}}=0.$$
The odd integer $N$ is to be determined, together with the remaining grid points along the latitude, $\beta_i$ with $i=1\ldots N$. 
To do this we need to find the costant $K < 1$ such that,
given
\begin{equation}
\beta_i=\beta_{i-1}-\frac{K}{N_D \sin(\beta_{i-1})}
\end{equation}
we get $\beta_{i^*}=0$, having started to iterate from $i=N$.
Once we have found $i^*$ we can get $N=2 i^*-1$. 
Hence, found all the $\beta_i$, we can find 
\begin{equation}
\Delta \lambda_i=\frac{H}{N_D \cos\beta}
\end{equation}
where $H<1$ is the maximum value such that $\Delta \lambda_i$ is a submultiple of $2 \pi$.

\section{Robust statistic}
\label{ROBUST}
The presence of disturbances in real data, both in time and frequency domain, 
leads to the need to use robust veto criteria, as discussed in Sec.\ref{pmclean}. To this purpose the use of statistical procedures based on
the median of the population, rather than on the mean, is often useful being the median much more robust with respect to the presence of tails in the distribution of a given random variable.
We have used the median to construct a robust estimator of the dispersion parameter, and used
it instead of the classical standard deviation.
The robust statistic consists in describing the statistical properties of a random variable $x$ through the median $m(1)=median(x)$ and a dispersion parameter defined as:

\begin{equation}
m(2)=\frac{median(abs(x)-m(1))}{c}
\end{equation}

where $c=0.6745$ is a normalization factor such that, if the distribution of $x$ is normal, then $m(2)$ is the standard deviation.

\bibliography{HOUGHmain}

\end{document}